\documentstyle[11pt,aaspp4]{article}
\def\A{\,\AA\,}
\lefthead{Ogle et al.}
\righthead{Polarization of BAL QSOs}

\begin{document}
\title{Polarization of Broad Absorption Line QSOs \\
       I. A Spectropolarimetric Atlas}

\author{P. M. Ogle \altaffilmark{1}, 
        M. H. Cohen \altaffilmark{2},
        J. S. Miller \altaffilmark{3}, \\
        H. D. Tran \altaffilmark{4}, 
        R. W. Goodrich \altaffilmark{5},
        and A. R. Martel \altaffilmark{6}} 

\altaffiltext{1}{Center for Space Research, Massachusetts Institute 
        of Technology, NE80-6095, 77 Massachusetts Avenue, Cambridge, 
        MA  02139, pmo@space.mit.edu}
\altaffiltext{2}{Mail Stop 105-24, California Institute of Technology, 
        Pasadena, CA 91125}
\altaffiltext{3}{UCO/Lick Observatory, University of California, Santa Cruz, 
       CA 95064}
\altaffiltext{4}{Institute of Geophysics and Planetary Physics, Lawrence 
       Livermore National Laboratory, Livermore, CA 94550}
\altaffiltext{5}{The W.M. Keck Observatory, 65-1120 Mamalahoa Highway, 
       Kamuela, HI 96743}
\altaffiltext{6}{Space Telescope Science Institute, 3700 San Martin Drive, 
       Baltimore, MD 21218}

\begin{abstract}

We present a spectropolarimetric survey of 36 broad absorption line 
quasi-stellar objects (BAL QSOs). The continuum, absorption trough, and 
emission line polarization of BAL QSOs yield clues about their structure. 
We confirm that BAL QSOs are in general more highly polarized than 
non-BAL QSOs, consistent with a more equatorial viewing direction for the 
former than the latter. We have identified two new highly-polarized QSOs in 
our sample (1232+1325 and 1333+2840). The polarization rises weakly to the 
blue in most objects, perhaps due to scattering and absorption by dust 
particles. We find that a polarization increase in the BAL troughs is a 
general property of polarized BAL QSOs, indicating an excess of scattered 
light relative to direct light, and consistent with the unification of 
BAL QSOs and non-BAL QSOs. We have also discovered evidence of 
resonantly scattered photons in the red wing of the C IV broad emission lines
of a few objects. In most cases, the broad emission lines have lower 
polarization and a different position angle than the continuum.
The polarization characteristics of low-ionization BAL QSOs are similar
to those of high-ionization BAL QSOs, suggesting a similar BAL wind geometry.
\end{abstract}


\clearpage
\section{Introduction}

Many galaxies contain active galactic nuclei (AGN), thought to be powered by 
accretion onto super-massive ($M \simeq 10^8 M_{\odot}$) black holes. The most
powerful AGN are the quasi-stellar objects (QSOs), with luminosities in the 
range $10^{44}-10^{48}$ erg/s. They out-shine their host galaxies by factors 
of up to $10^4$. These extremely luminous objects provide laboratories to 
study high-energy physics and also may play a crucial role in the formation 
and evolution of galaxies. 

About 10\% (\cite{wmf91}) of all QSOs have broad absorption lines (BALs) in 
their spectra indicating a fast (up to 0.1~c or greater) outflow. These are 
the broad absorption line quasi-stellar objects (BAL QSOs). ~\cite{ssh97} list 
304 QSOs (plus 133 candidates) with associated absorption lines or BALs. The 
BAL phenomenon appears primarily in high-luminosity radio-quiet objects. BAL 
do not appear in Seyfert galaxies and have only recently been discovered in a 
few radio-loud QSOs (\cite{bgh97}). However, low-velocity intrinsic 
absorption systems are seen in all subtypes of broad-line AGN, indicating that
outflows are a common phenomenon.

A survey of 142 bright QSOs (\cite{sma84}) shows that the great majority 
(99\%) have low polarization (LPQ, $P<3\%$). BAL QSOs make up a large
fraction of the radio-quiet highly polarized QSOs (HPQ, $P>3\%$). There is a 
strong selection bias against reddened and obscured QSOs in optical
surveys. Some infrared-selected AGN have been discovered which would be 
classified as QSOs if seen from a dust-free direction (e.g.,~\cite{wwe92}; 
~\cite{wh97}; and ~\cite{gmm96}). These objects are very highly polarized 
(up to 30\%) because the unpolarized direct emission is more highly attenuated
than the polarized scattered emission. The scattering region is located 
further away from the nucleus than the broad line region, as indicated by 
polarized broad emission lines.

The radio-loud HPQs are mostly blazars, which emit optical synchrotron light
boosted by relativistic motion in a jet pointed close to the line of sight. 
Another class of highly polarized AGN consists of reddened quasars 
(e.g.,~\cite{gm88};~\cite{rs88};~\cite{bwd98};~\cite{dbt98}) and radio 
galaxies with hidden quasar nuclei (e.g.,~\cite{tcg95};~\cite{ocm97};
~\cite{tco98};~\cite{cot99}).
 
The question of whether BAL QSOs can be unified with other types of QSOs has 
important implications for the structure and physics of AGN outflows. It has 
been suggested that BAL QSOs are viewed from an equatorial direction, while 
non-BAL QSOs are viewed from a more polar aspect (\cite{wmf91}). In this 
picture, the BAL clouds may be ablated from an accretion disk. An equatorial 
aspect is consistent with the higher mean polarization of BAL QSOs than 
non-BAL QSOs (e.g.,~\cite{sh99}). Our survey of BAL QSOs improves the 
accuracy and statistics of the BAL QSO polarization distribution. Ogle 
(1999b, hereafter Paper II) tests the hypothesis that high polarization is 
due to viewing angle by comparing the polarization distributions of BAL QSOs 
and non-BAL QSOs against a simple unification model. 

The source of polarized light in low-polarization QSOs is a subject of 
debate (e.g., ~\cite{wms93}). UV-optical emission from QSOs (the big blue 
bump) is commonly thought to be dominated by thermal emission from an 
accretion disk (e.g., ~\cite{sm89}). It was originally predicted that 
such emission should be highly polarized (up to 12\%) due to scattering 
(\cite{ls75};~\cite{c60}). The polarization should also be perpendicular to 
the disk axis for an optically thick disk. The observations, however, show 
that the polarizations of QSOs are low and sometimes parallel to the radio 
axis, weighing heavily against the standard accretion disk model (e.g.,
~\cite{agg96}). In Paper II, we consider scattering in the broad emission line
region as a possible source of polarization. 

The UV spectral components of QSOs include the blue featureless continuum (FC),
broad emission lines (BELs), broad absorption lines (BALs), and sometimes 
narrow absorption lines (NALs). Most of the variance among QSO spectra can be 
associated with these components (\cite{fhf92}). Identifying and measuring 
individual lines and defining a continuum level can be difficult because of 
blending, especially of Fe II emission (see, e.g.,~\cite{wnw85}). 
Spectropolarimetry provides a way of disentangling distinctly polarized 
spectral components . 

Spectropolarimetry has been published for a small number of BAL QSOs (e.g., 
~\cite{sah81};~\cite{gsf94};~\cite{gm95};~\cite{cot95};~\cite{hw95};
~\cite{btb97};~\cite{sh99}). In general, these studies reveal that the broad 
emission lines have lower polarization than the continuum and that the 
continuum polarization rises to the blue. Our survey extends these results to 
a much larger sample of objects. In addition, the large apertures of the
Keck Telescopes allow us to study the emission and absorption line
polarization of distant QSOs with unprecedented detail. This is particularly 
difficult in the deep BAL troughs, where the equivalent apparent magnitude
is typically 20, and we are trying to measure the polarization to an accuracy
of 1\%. As a byproduct, we have taken some of the highest S/N spectra of
BAL QSO to date. The authors will gladly provide total flux spectra in FITS 
format to those who can use them in their research. 
(Contact P. Ogle at pmo@space.mit.edu.)

This is the first in a series of three papers on the polarization of BAL QSOs.
In \S 2 we discuss the sample selection. In \S 3 and \S 4 we describe the Keck 
and Palomar observations. Data reductions and calibration procedures are 
presented in \S 5. In \S 6 we tabulate the results of the spectropolarimetric 
survey, including continuum and BAL polarization. The general polarization
characteristics of BAL QSOs are discussed in \S 7. The spectra of individual 
objects are presented and discussed in \S 8. Paper II (\cite{o99b}) covers
the polarization distribution of BAL QSOs and the wavelength dependence of the
polarization. Paper III (\cite{o99c}) is a detailed study of the broad 
absorption line and broad emission line polarization of BAL QSOs. For a
preview of Papers I-III, see also Ogle 1998. 

\section{Sample Selection}

A sample of 24 BAL QSOs was observed with the Double Spectrograph 
polarimeter on the 5-m Hale telescope to determine their polarization spectra.
Of these, 21 were selected from the Weymann et al. (1991, WMF) sample of 40 
BAL QSOs. This was augmented by 3 other BAL QSOs fitting into gaps in the 
observing schedule. Observations were restricted to declinations 
$\delta>-30\arcdeg$. The WMF sample was derived from the LBQS and a number of 
other surveys. This sample is useful because it contains most of the 
well-studied bright BAL QSOs accessible from northern latitudes. However, due 
to haphazard selection and serendipitous discovery of most BAL QSOs, the sample
should not be considered uniform or complete.  

In addition, we observed 21 BAL QSOs with the W. M. Keck 10 m telescopes. The 
Keck survey provided the high signal-to-noise ratio (S/N) necessary to study 
BAL and BEL polarization. A total of 9 BAL QSOs from the Palomar survey were 
re-observed at Keck. The remaining 12 BAL QSOs were selected to be bright or 
have high polarization. In cases of overlap between the two surveys, we
present the superior Keck data. The combined data set consists of Keck 
spectropolarimetry of 21 BAL QSOs and Palomar spectropolarimetry of 15 BAL 
QSOs.

We give a summary of the observations in Tables 1 and 2, including IAU and 
alternate designations, redshift, apparent visual magnitude, UT date, exposure
time, and spectrograph slit position angle. Total exposure times range from 
0.5 to 6.7 hrs, for an average of 2.2 hrs per object. Our results for 
0105$-$265 and PHL 5200 were first presented by Cohen et al. (1995); we 
discuss them in additional detail.  All objects have galactic latitudes 
$|b|>30\arcdeg$, so interstellar polarization is typically less than 0.2\%. 
(See Appendix B for a table of estimated interstellar polarizations.) The 
apparent visual magnitudes range from V=16.5-19. 

\section{Keck Observations}

Spectropolarimetry was taken with the Low Resolution Imaging Spectrograph 
(LRIS, ~\cite{o95}) and polarimeter combination. A 300 groove/mm 
grating with a dispersion of 2.49\A /pixel gave a resolution of 10\A. This 
translates into a velocity resolution of 600 km/s in the C IV $\lambda$1549
BAL for a QSO at redshift $z=2.2$. The C IV doublet separation of 500
km/s is unresolved in our spectra. This resolution is sufficient for studying
the polarization of the broad emission and broad absorption lines. 

The spectra cover the observed wavelength range of 3800-8900\A. This 
corresponds to the rest wavelength range between Ly$\alpha$ $\lambda$1216 and 
Mg II $\lambda$2798 for a $z=2.2$ QSO. The C IV $\lambda 1549$ line is the most
useful for our study because this BAL is typically deep enough to show a large
polarization signature and it is well isolated from other emission
and absorption lines, including intervening Ly$\alpha$ absorption systems.
There are Mg II and Fe II metal lines from intervening systems in the spectra
of at least ten objects; but their equivalent widths are much smaller than the 
BAL equivalent widths. The C IV BAL is not visible in the two objects with 
redshifts $z<1.6$.

The detector is a 2048$\times$2048 pixel charge coupled device (CCD) 
manufactured by Tektronix. It has an inverse gain of 2 electrons per data 
number and a read-out noise of roughly 6 electrons. The maximum 
efficiency of the spectrograph plus CCD combination is approximately 40\% at 
the blaze wavelength of the grating. 

Slit widths of $1\arcsec$ and $1\farcs 5$ were used to match 
the atmospheric seeing, and the pixel scale was $0\farcs2$/pixel. 
The useful slit length is only $30\arcsec$ in LRIS polarimetry mode 
because of vignetting by the beam-splitter. We assume that the extended 
emission from the narrow line region and starlight from the host galaxy within
the slit aperture are negligible. There are a few cases where foreground 
galaxies fell in the slit, but they were never close enough to the BAL QSO to 
contaminate its spectrum (except for the undetected galaxy which lenses 
1413+1143). Indeed, there is no evidence for narrow emission lines or stellar 
photospheric absorption lines in any of our spectra.

The relatively narrow slit widths along with variable weather and seeing 
conditions yield large uncertainties in the absolute photometry, however we 
are only interested in relative fluxes. The slit orientation on the sky was 
held constant during each set of observations to avoid mixing the Stokes 
parameters. When observations from different nights were taken with different 
slit PA, the Stokes parameters were averaged after calibration to the sky 
coordinate system. The spectrograph slit was set at the mean parallactic angle 
for most observations. This resulted in a minimum loss of light from 
differential atmospheric refraction, but there was still up to 20\% 
differential loss between the blue and red ends of the spectra. This is seen 
in the ratio of spectra taken at different epochs. It is important to consider
this distortion when analyzing the color of the flux spectra, but it does not 
affect the polarization.

Keck BAL QSO observations were taken without an order-blocking filter to
obtain a wide wavelength coverage, and may be contaminated by second-order 
blue light redward of 7000\A. The magnitude of this effect (\cite{pc96}) is 
less than 1.8\% of the flux at the blue end of the spectrum (3500-4500\A). 
Contamination due to second order light is minimal, since the BAL QSO continua
rise less than a factor of 3 over the observed wavelength range. This 
translates to an error of 5\% in flux at the red end of the spectrum of the 
bluest object. Since the polarization in BAL QSOs rises to the blue, this 
could lead to a polarization error of $\delta_P=0.08\%$ in an object polarized
at $P=2\%$. There is a potentially larger effect from the Ly$\alpha$ emission 
line showing up in second order at 2432\A, but we have not identified any 
instance of this in our spectra.

The performance of the LRIS polarimeter and the robustness of the reduction 
techniques are discussed by Cohen et al. (1997). The dual-beam polarimeter 
separates and simultaneously measures two orthogonal polarization senses using
a calcite beam splitter. This nearly eliminates the effects of atmospheric 
transparency and seeing variations on the calculated Stokes parameters.
Rotating a half-wave plate $45\arcdeg$ swaps the ordinary and extraordinary 
beams, allowing a correction for the gain difference along the two optical 
paths. A full set of observations is taken at four wave plate settings  
(0$\arcdeg$, 45$\arcdeg$, 22\fdg 5, 67\fdg 5), yielding eight spectra used to 
compute the Stokes parameters ($F$, $Q$, $U$). The circular polarization is 
negligible for QSOs. 

\section{Palomar Observations}

The Palomar observations are summarized in Table 2.
Spectropolarimetry was taken with the Double Spectrograph (\cite{og82}) on 
the 5-m Hale Telescope. BAL QSOs were observed for 60-90 minutes, depending on 
their magnitudes and polarizations. Polarization spectra are measured with an 
accuracy of 0.1-0.6\% in 50\A (rest) bins. The methodology of observations is
largely the same for the Palomar and Keck samples. Here we concentrate on 
issues peculiar to the Palomar data. 

We used a D55 dichroic filter, splitting the spectrum at 5500\A. Separate 
gratings were used to disperse the spectra in the blue and red arms of the 
spectrograph. The transmittance of the dichroic is better than 90\% redward 
of 5500\A, so only a small fraction of the photons is lost. The dichroic 
filter has the additional advantage of blocking out second-order blue light.
The polarimeter is above the dichroic in the optical path, so it serves both 
the blue and red arms of the spectrograph. Some spurious polarization effects 
are introduced near the wavelength of the dichroic split and do not 
calibrate out. We set the gratings on the blue and red sides to exclude 
this region, so there is typically a 200\A gap in our spectra. For most 
objects, this gap falls in between the C IV and C III] BELs. However, the gap 
falls in the C IV BAL region for three objects (0903+1734, 1235+1453, and 
1243+0121), and on the C III] emission line for two objects (0932+5006, and 
2201$-$1834). 

A 300 line/mm grating blazed for maximum efficiency at 3990\A gives a 
dispersion of 2.17\A per pixel at the focus of the blue camera. A 316 line/mm 
grating blazed at 7500 \A gives a dispersion of 2.46\A per pixel in the red
spectra. The wavelength coverage is typically 3600-5400\A in the blue and 
5600-8100\A in the red. The low efficiency of the Double Spectrograph at 
wavelengths less than 3600\A is not adequate for polarimetric observations of 
faint QSOs. The spectrograph slit had a $2\arcsec$ aperture and a 
useful (un-vignetted) length of $35\arcsec$ in polarimetry mode. We 
used a wider slit at Palomar than at Keck because of the worse seeing. The 
spectral resolution was 6\A in the red and 8\A in the blue, slightly better 
than in the Keck observations.

The detectors in the blue and red cameras were both 800$\times$800 pixel CCDs.
The blue detector suffered from dead columns and transient gain variations 
near the blue end of the spectrum. We used chip regions with the best cosmetic 
quality, but the flux spectra may still be afflicted by nonlinearities at a 
level of 5\% in localized wavelength regions. In fact, the flux standards of 
Oke (1990) were taken with the Double Spectrograph and suffer from some of the
same defects. These glitches in the system efficiency were at a low enough 
level to cause negligible damage to the polarization spectra. 

The polarimeter in the Double Spectrograph does not have focus compensation 
between the ordinary and extraordinary beams, so the camera foci were set at 
the best compromise positions. The red camera CCD chip suffered from a warped 
surface, causing focus variations along the spectra. We set the telescope
focus to provide the sharpest image along the largest possible wavelength 
range. These focusing difficulties should not affect the polarimetry (except 
for lowering the S/N) since focusing and  seeing variations are mitigated by 
the dual-beam polarimetry method. During the last year of our observing 
program (1995-1996), the red camera was upgraded to a 1024$\times$1024 pixel 
CCD and other changes were made to its optics, removing the problems of 
multiple detector flaws and some of the focusing difficulties.

\section{Data Reduction and Calibration}

Data reduction was accomplished with VISTA 
\footnote{VISTA was originally developed by R. J. Stover and T. R. Lauer in 
1982. It is primarily used at Lick Observatory. The current version is 
maintained by John Holtzman (email holtz@nmsu.edu).} 
image processing software, using procedures described by Miller, Robinson, 
\& Goodrich (1988) and Cohen et al. (1997). Cosmic ray events were eliminated 
interactively from the CCD frames using a $3\sigma$, $5 \times 5$ pixel median 
filter. It was sometimes impossible to remove cosmic ray tracks falling on 
the object spectra, and these are flagged in the spectra as 'hits'. To help 
identify and mitigate contamination from cosmic rays, the observations were 
typically split into two sets. The CCD frames were de-biased using an average 
from the over-scan region. No corrections were made for the negligible dark 
current. All object frames were divided by an internal halogen flat 
field to remove small spatial variations in the CCD response. Note that in the
case of low sky background, the flat field cancels out of the equations 
for the Stokes parameters. Hence the polarization is quite insensitive to 
flat-fielding. 

The object spectra were extracted using 3-$6\arcsec$ wide windows 
centered along a spline fit to their centroids on the CCD chip. Night sky 
spectra were extracted from windows on both sides of the object, averaged, 
scaled, and subtracted from the object spectra. Spectra were wavelength 
calibrated using Ne, Ar, and Hg+Kr internal lamps, and re-binned to a linear 
scale. Night sky spectra were used to correct for wavelength zero-point 
offsets due to spectrograph flexure. Spectra were de-redshifted using 
literature redshift values given in Tables 1 \& 2 and binned by 5\A (rest). 
The photon counts and statistics from the object and sky spectra were then 
used to compute the Stokes parameters ($F$, $Q$, $U$) and their uncertainties. 
Binning reduces the biases that can be introduced by noisy data and enhances 
the signal-to-noise ratio S/N of the Stokes parameters. The total flux spectra
(Figs. 3-5) are presented un-binned to show maximum spectral resolution.

Measurements of ($Q$, $U$) and the derived fractional polarization and 
position angle ($P$, PA) are subject to a number of biases when the number 
of photons is small (\cite{cs83};~\cite{nc93}). $P$ is a positive definite 
quantity, and therefore does not obey Gaussian statistics. It is therefore 
preferable to rotate ($Q$, $U$) by a fit to the continuum PA curve
instead of calculating a de-biased $P$ (\cite{ss85}). This is only possible 
when the PA is a slowly varying function of wavelength. Where there are PA 
rotations across the emission and absorption lines, the polarization will
be underestimated by this method; then it is necessary to inspect both
$Q$ and $U$ to determine the line polarization. 

Flux spectra were corrected for atmospheric extinction (\cite{h71};
~\cite{bbd88}) and absorption in the telluric bands ($\rm{O}_2$ A, B, and 
$\gamma$, and $\rm{H_2O}$). Absolute flux calibrations were made using 
spectrophotometric standards (\cite{o90}), but some nights were not 
photometric. Flux standards were observed at Keck both with and without a 
GG-495 order-blocking filter to eliminate contamination by second-order light.
 
Bright stars were observed through UV and IR polarizers to determine the 
position angle of the Keck and Palomar half-wave plate fast axes as 
functions of wavelength. The half-wave plates are super-achromatic 
(\cite{g91}), designed to give a retardance close to $180\arcdeg$ over a large
wavelength range. However, this design introduces a wavelength-dependent 
rotation into the instrumental PA curves. The Keck PA correction curve 
(Fig. 1) was measured on several occasions and averaged to 
obtain a curve with relative uncertainty of $<0\fdg 2$ at all wavelengths. 
All Keck observations were corrected by a spline fit to this average curve. 
A similar PA correction curve was measured for Palomar data.

The PA offset between the half-wave plate coordinate system and the sky 
coordinate system was determined for each observing run using polarized 
standard stars from the list of Schmidt et al. (1992). $B$ and $V$-band 
polarizations of the standards were measured by averaging over the bands 
3950-4900\A and 5050-5950\A, respectively. Some of the standards show 
curvature in the PA spectrum, amounting to 1-$2\arcdeg$ over the range 
3800-8900\A. The PA variation within the B or V band is less than $0\fdg 5$,
so the error introduced by not considering the shape of the Johnson 
filter curves is less than $0\fdg 2$. We used a set of secondary and 
tertiary PA standards (Table 3) dim enough to view with the Keck Telescope and
covering the sky. The PA values of the tertiary standards were set by 
minimizing the difference between PA offsets determined from all secondary 
standards over a number of observing runs. The absolute PA calibration is 
internally consistent to $\pm 0\fdg 6$. This is commensurate with the random 
uncertainties, which are as large as $0\fdg 4$ for the published secondary 
standards. 

Null polarization standards (\cite{sel92}) were observed on several 
occasions at Keck (Table 4), and were always found to be null 
to within 0.1\% (4000-7000\A). The standard deviation of all null observations
was ($\sigma_Q$, $\sigma_U$)$=(0.05\%,0.05\%)$. There was no evidence for 
systematic instrumental polarization effects, so no instrumental polarization 
correction was required for Keck data. The lack of instrumental polarization 
indicates near-perfect alignment and uniform reflectivity of the primary and 
secondary telescope mirrors. Before PA correction, the Palomar data were 
corrected for significant instrumental polarization (see Appendix A). We 
discuss and tabulate estimates of the interstellar polarization in Appendix B.

\section{Continuum and Broad Absorption Line Polarization}

Table 5 gives the continuum polarizations for the Keck BAL QSO
sample. We measured the continuum polarization in two narrow wavelength bands,
$P_1$(1600-1840\A, rest) and $P_2$(1960-2200\A, rest). These bands are on 
either side of the C III] $\lambda 1909$ emission line, where the 
contamination from Fe II and other emission lines is relatively low. 
Broad-band measurements $P_c$ and PA (4000-8600\A, observed) are presented for 
comparison with polarization values in the literature. The polarization $P$ 
was estimated by rotating the Stokes parameters by the angle 2PA, 
determined from a cubic spline fit to the PA curve. (The Stokes parameters 
become $(Q^{\prime}=P, U^{\prime}=0)$ in the rotated coordinate system.) $P$ 
was then averaged over the appropriate bins in the spectrum. This introduced 
negligible bias in the high S/N continuum observations. Objects with 
significant PA rotations in the continuum are indicated by an 'r' in the 
last column. Polarization values are given as percentages and PA values are 
given in degrees.

We measured the continuum polarization of the Palomar BAL QSOs in two continuum
bands (1600-1840\A and 1960-2260\A). For most objects, only one of these two
bands was available and the other fell in the dichroic gap. Except for a few 
observations heavily affected by clouds, we measured continuum polarization 
with an uncertainty of $<0.3\%$. This approaches the 
level of the systematic uncertainty in the instrumental polarization curve
(see Appendix B). The results of our continuum polarization measurements at 
Palomar are presented in Tables 6 \& 7. Several objects (9/24) 
are rated as polarization non-detections, and have $Q/\sigma_Q<3$ and 
$U/\sigma_U<3$. We list $Q$ and $U$ only for these objects, since $P$ and PA
are biased at low S/N.

We present a comparison of polarization and PA values for objects measured
both at Palomar and Keck in Figure 2. Except for 1235+0857,
all of the measurement pairs are within $2\sigma$ of the equality line. There 
is no indication of gross systematics or strong variability between the two 
sets. We discuss the weak continuum polarization variability, which may help 
constrain the source of polarized continuum radiation, in Paper II.

Broad absorption line (BAL) polarization values are listed in Table 8. The 
velocity, rotated Stokes parameters, and PA rotations were determined at the
point in the trough where $Q^{\prime}$ is a maximum. The trough polarization 
values and uncertainties are for 5\A (rest) bins, which are roughly 1.5 
resolution elements wide for a $z=2$ QSO. Binning inevitably reduces the peak 
trough polarization, but provides a more reliable value because of reduced 
noise (see Paper III). 

Small but significant ($>3\sigma$) PA rotations are seen in the BAL troughs 
of 0226$-$1024, 07598+6508, 0903+1734, 1246$-$0542, and 1524+5147. PA 
rotations are also seen in the C IV troughs of 0146+0142, 1212+1445, 
1333+2840, and 1413+1143, after averaging over several bins. PA rotations in 
the BALs may indicate an asymmetry in coverage of the continuum scattering 
region by the BAL clouds. In general, the peaks in $Q^{\prime}$ and 
$U^{\prime}$ can be at different trough velocities, but there is a tendency 
for the polarization to peak where the trough is the deepest. The error made 
in estimating the trough $P$ from $Q^{\prime}$ is less than 0.2\% for most 
objects because the PA rotations are small. In objects where the BAL PA 
rotates by $>10\arcdeg$, both $Q^{\prime}$ and $U^{\prime}$ must be inspected 
for a full description of the trough polarization. 

\section{General Results and Discussion}

The polarization of BAL QSOs is generally stronger than that of non-BAL QSOs,
but is weak in absolute terms. Most of the BAL QSOs in our sample have low to 
moderate continuum polarization ($P_c=0-3\%$). We have identified two new 
high-polarization QSOs (HPQ) in our sample (1232+1325, $P_1=3.38\pm 0.08\%$; 
and 1333+2840 (RS 23), $P_2=5.61 \pm 0.07\%$). There are now six known 
high-polarization BAL QSOs (RS 23, PHL 5200, CSO 755, 1232+1325; 
FIRST 0840+3633 and FIRST 1556+3517, ~\cite{btb97}). The polarization 
distribution is discussed in detail in Paper II, and tested against a QSO 
unification model. We come to the conclusion that the polarization 
distributions of BAL QSOs and non-BAL QSOs are consistent with a BAL outflow
at equatorial or intermediate latitudes. 

We show in Paper II that there is no simple correlation between BEL or BAL 
equivalent widths and continuum polarization. Goodrich (1997) 
hypothesizes that the polarization in HPQ may be enhanced by attenuation
of the diluting central continuum source. He notes that the HPQ tend to
have large emission line equivalent widths (C III]+Al III rest equivalent 
width REW$>42$\AA). This may further indicate that an attenuating medium 
preferentially blocks the continuum source but not the BEL region. He 
predicts that BAL QSO with large REW BELs should be highly polarized. While we
find this to be the case for 1232+1325 (REW$=51$\AA) and RS 23 
(REW$=45$\AA), it is not so for 1235+1453 (REW$=59$\AA) or 1243+0121 
(REW$=64$\AA). The HPQ fraction is higher in high-REW BAL QSO than in low-REW 
BAL QSO, but the statistics are poor. Even though there is no direct
relationship between $P$ and BEL REW, attenuation may still be important in 
some highly polarized BAL QSOs.

The wavelength dependence of the continuum polarization in BAL QSOs is
weak. A few objects show variation of continuum PA with wavelength, indicating
the presence of at least two polarization sources. The largest continuum PA 
rotations are seen in UM 232, 0145+0416, 0932+5006, and 1700+5153.
While interstellar polarization may be important in some of these cases, the
wavelength independence of the PA, unpolarized emission lines, and the high 
galactic latitude of most objects suggests there is little contamination by 
interstellar dust (see Appendix B). Most BAL QSOs (14/21 objects observed at 
Keck) show a rise in polarization to the blue. Nine BAL QSOs show the 
polarization dilution signature of broad Fe II line emission between 
2260\A ~and 2800\A. This contributes to but doesn't completely account for the
wavelength dependence of the polarization (Paper II). Of the remaining 
objects, 5 have flat polarization spectra, one (UM 232) has polarization 
rising to the red, and one object is unpolarized (UM 275). The wavelength 
dependence of the polarization is most likely caused by dust scattering and 
absorption intrinsic to the QSOs. We consider the detailed implications of 
this in Paper II.  

The peak C IV BAL polarization is at least $2\sigma$ greater than the 
continuum polarization for most objects (17/20), and can reach values as
large as 17\%. There are also PA rotations across the BALs of 9 objects. The 
polarization variations across the troughs can be attributed to polarized 
light scattered around the BAL region. The scattered lines of sight on average
pass through a lower BAL optical depth, and the BALs are shallower in 
polarized flux than in total flux. This is yet another indication that the BAL
region does not uniformly cover the central source, and that a BAL QSO may 
appear as a non-BAL QSO from a different viewing direction. In some objects, 
the BALs are blue-shifted in polarized flux, relative to their velocities in 
the total flux spectra. This may tell us something about the dynamics of the 
BAL outflow, and is discussed in detail in Paper III. 

While the dominant BAL polarization effect is from partial coverage of 
the continuum scattering region, there are cases where a small amount of 
resonantly scattered light from the BAL region is also detected (Paper III,
and \cite{o97}). This is important because it is the first time the expected 
resonance scattering profile (\cite{hkm93}) is seen. It is proof that at 
least some photons resonantly scattered out of the line of sight by the BAL 
clouds escape the QSO. Higher S/N observations of the resonance scattering
profile can potentially be used to constrain the geometry and dynamics of the 
BAL outflow. In some objects the PA in the high-velocity 
($\sim 10,000$ km/s) red wing of the C IV BEL rotates in the opposite 
direction from the BAL PA rotation. Polarized flux at high velocities 
with a profile distinct from the BEL profile is a good indication of resonance
back-scattering. Photons from the central source are reflected forward by the 
portion of the BAL wind flowing away from the observer. There are indications 
that C IV BEL photons are also back-scattered by the BAL (Paper III), but at a 
different PA. The location and nature of the scatterers is discussed in 
detail in Papers II and III.

In general, the BELs are polarized at a lower level than the continuum 
and to first order do not show up in polarized flux. A closer look 
shows small rotations and residual polarized flux across some emission lines 
in some objects (Paper III). Most notably, C III] is polarized in 2225$-$0534
and 0019+0107. Weakly polarized C IV photons are also detected in a few 
objects, as discussed above. The line polarization is used as a diagnostic of 
the geometry of the BAL, BEL, and scattering regions in Paper III.

In addition to the 10 previously known low-ionization BAL QSOs in our
sample, we have found weak low-ionization BALs or associated NALs
(Al III and/or Mg II) in 9 of the objects with high S/N Keck observations. 
More than half (52\%) of the objects in our sample have intrinsic absorption
from low-ionization species. The polarization characteristics of the objects 
with strong low-ionization BALs (and even the Fe II Lo-BAL QSO 0059-2735) are 
similar to those with weak or non-existent low-ionization BALs. We 
therefore argue that there is no great difference in the geometry of the
BAL outflow in high and low-ionization BAL QSOs. 

\section{Individual objects}

We discuss individual polarized BAL QSOs (and two interesting unpolarized BAL
QSOs) in this section. The spectra of 29 BAL QSOs are presented in Figure 3. 
There are two objects per page, and four panels per object. The four panels 
show the flux $F_{\lambda}$, polarization $P=Q^{\prime}$, polarized flux 
$Q^{\prime}\times F$, and position angle $\theta$. When viewing 
polarimetry data, it is useful to think of the total flux (panel 1) as mainly 
due to a direct, unpolarized component plus a small fraction of scattered 
light. The scattered light spectrum is well represented by the polarized flux 
(panel 3). The scaling between the total scattered flux and the polarized flux
depends on the polarizing efficiency and optical depth of scatterers. The 
polarization (panel 2) is just the ratio of panels 3 and 1, so does not 
contain any independent information. Short descriptions of the spectra are 
given below. Selected objects are discussed in more detail in Papers II and 
III. Palomar spectra (Figure 4) typically reach further into the blue than the
Keck flux spectra. For example, the Palomar spectrum of 0059$-$2735 shows the 
entire C IV BAL, inaccessible at Keck. The flux spectra of the remaining 
objects with no detectable polarization are displayed in Figure 5.

1. {\it 0019+0107 (UM 232).} The C IV and Si IV BALs show multiple narrow
sub-components. There is an intervening metal line absorber. The continuum 
polarization drops rapidly blueward of 2000\A. This is unique to UM 232 since 
all other objects have $P$ flat or rising to the blue. There is a PA 
rotation across the continuum corresponding to the drop in polarization. This 
strongly suggests at least two sources of polarized light. We consider the 
possibility of interstellar contamination in Paper II. There is a 
characteristic rise in polarization in the C IV trough, in spite of the 
rapidly falling continuum polarization at the position of the line. The C III]
$\lambda1909$ emission line shows up strongly in polarized flux. Polarized 
C IV emission is weakly present (a 4$\sigma$ detection). Polarized C III]
is also seen in PHL 5200 (see below), and may be due to resonance scattering 
in the BEL region. 

2. {\it 0025$-$0151 (UM 245).} This object has very shallow BAL troughs, 
punctuated by narrow absorption components. The C IV BAL changed 
dramatically between 11/95 and 10/96 (Fig. 3 shows the 10/96 state). It 
deepened and 3 or 4 new narrow BAL components appeared. The presence of 
variability in such a weak BAL suggests that weak BALs may be missed in some 
QSOs, depending on the BAL equivalent width in the discovery spectra. This  
would decrease the apparent BAL QSO fraction. The polarization is low, and may
increase to the blue. The PA spectrum is quite noisy and the apparent 
rotation in the blue is not statistically significant. 

3. {\it 0043+0048 (UM 275).} This object is remarkable for its deep double 
troughs. The continuum polarization is null, with a formal $2\sigma$ upper 
limit of 0.2\%. The broad emission lines have unusually narrow profiles and 
are unpolarized. Low polarization indicates a high degree of symmetry in 
the continuum emitting region and a low optical depth to scattering throughout
the nuclear regions. The BAL troughs show significant polarization 
($Q=4.4\pm0.9\%$ at -3700 km/s). This measurement is difficult because the 
troughs are deep, and needs to be confirmed. The BAL polarization may be 
entirely due to resonance scattering since the continuum and emission lines
are unpolarized. The C IV troughs are unusually deep for a BAL QSO, but 
residual flux appears at the bottom of the troughs. There are Mg II and Al III
BALs, also consistent with a large column density of BAL plasma. 
 
4. {\it 0059$-$2735.} This was the first object discovered in the rare class 
of Fe II low-ionization BAL QSOs (\cite{hmw87}). The polarization increases in
the Mg II and Al III BALs. There are possible PA rotations in these BALs but
they have low significance. The similar polarization behaviors in the BALs of 
low-ionization and high-ionization BAL QSOs suggest that they share a similar 
geometry. The appearance of low-ionization BALs may require a total hydrogen 
column $N_H>10^{22}\rm{cm}^{-2}$ (\cite{h99}). This is expected of lines of 
sight passing near the QSO accretion disk. Most BELs, including Fe II and 
Mg II have low polarization, similar to those of high-ionization BAL QSOs. 

In contrast to the BAL polarization, the polarization decreases in the narrow 
absorption lines (NAL), including Fe II, Zn II, Cr II, and Ni II 
(see ~\cite{wcp95} for a high-resolution spectrum and line identifications). 
As a result, the NAL appear deeper in polarized flux than in total flux. This 
is especially apparent in the Fe II NAL redward of C IV.  We attribute 
the low polarization in the NAL to dilution by low-polarization emission 
lines, especially Fe II. The NAL region must have a significantly different 
geometry from the BAL region. It appears that (for our line of sight) the NAL 
clouds cover the scattering region more completely, but the BEL region less 
completely, than do the BAL clouds (see Paper III for details). The low 
line-of-sight covering fraction of the Fe II BEL region by the NAL suggests 
that the NAL clouds are located close to and perhaps intermingled with the BEL
region. 

The Fe II NAL at 2380\A shows a $10\arcdeg$ rotation in PA, but the 
other Fe II NAL do not. This effect may be due to weak polarization of the 
Fe II emission lines filling the saturated NAL, and probably 
has no direct connection to the particular Fe II ionic transition. 
Perhaps the rotation is maximum at the wavelength of the Fe II $\lambda$2380 
NAL because this is where the Fe II emission line flux peaks. A similar PA 
rotation is seen in FIRST 0840+3633 (\cite{btb97}). 

5. {\it 0105$-$265.} This is the highest redshift object in our sample, with 
$z=3.488$, so it shows the largest number of absorption troughs. The 
spectropolarimetry was presented by Cohen et al (1995). The troughs are wide,
smooth and deep, similar to those of PHL 5200 (see below). The continuum 
polarization rises steadily to the blue and the PA is constant across the
entire spectrum, even in the troughs, blueward of Ly$\alpha$, and below the 
wavelength of the Lyman edge (912\A). The lack of a strong Lyman edge 
feature in direct or scattered flux suggests that both the BAL region and 
scattering region are highly ionized. 

There is a high contrast between the continuum and BAL polarization 
levels, with $P_c=1.62 \pm 0.07\%$ and $P_t \simeq 10\%$. $P$ increases in all
of the troughs, including C IV, Si IV, N V+Ly$\alpha$, Ly$\beta$+O VI and S VI.
Four of the troughs have nearly identical residual flux, suggesting they 
are saturated. However, the  peak trough polarization values do not follow the
trend of the continuum polarization, possibly due to  an ion-dependent 
line-of-sight covering fraction. The troughs are almost completely detached, 
so they are not contaminated by the BELs. 

6. {\it 0137$-$0153.} The BAL troughs have a complex morphology with several 
sub-components, including a deep one at low velocity. There is a weak
intervening metal absorption system. This BAL QSO has low continuum 
polarization increasing to the blue. $P$ rises in the C IV and 
Si IV BAL troughs, especially in the deepest sub-component. The polarization 
falls at all of the emission lines and they do not appear in the polarized 
flux spectrum. The spikes seen in the polarized flux and PA spectra at 
1250\A are probably due to a cosmic ray event.

7. {\it 0145+0416 (UM 139).} This object has BAL troughs similar in shape and 
depth to 1246$-$0542. There is a deep main trough superimposed on a 
broader trough of moderate depth. However, the main trough is less detached 
from the BEL in 0145+0416. The continuum polarization is moderate and may rise
to the blue. There is a strong continuum PA rotation from $126\arcdeg$ in the 
red to $98\arcdeg$ in the blue. The rotation starts blueward of the C III] BEL, 
similar to the rotations in 0019+0107 and 0932+5006. However, it is not 
accompanied by a red polarization spectrum like 0019+0107. A strong 
PA rotation requires at least two sources of polarized light with different 
PA and different spectral slopes. The maximum interstellar polarization of 
0.25\% could cause a PA rotation of $<2\fdg 4$. We conclude that
the PA rotation is due to multiple polarizing agents intrinsic to the QSO.

8. {\it 0146+0142 (UM 141).} This object is distinguished by high velocity 
absorption extending past 0.12c. The Si IV emission line appears to be 
completely occulted by the C IV BAL. There is a strong P V BAL (\cite{b93}),
and there may be a weak Al III BAL. There is also an intervening metal 
absorption system. The continuum polarization is wavelength independent. $P$ 
rises in the NV+Ly$\alpha$ BAL, the C IV BAL, and possibly the P V BAL. The 
spectrum is too noisy to determine the polarization of the O VI BAL. The 
PA rotates by $\sim 10\arcdeg$ in the Ly$\alpha$+N V, C IV, and Si IV troughs. 
$P$ decreases at the emission lines, but there is residual polarized emission 
line flux in O VI+Ly$\beta$, N V+Ly$\alpha$, and  C IV. However, the 
semi-forbidden C III] BEL does not appear in polarized flux. This difference 
between the polarization of the permitted and semi-forbidden BELs is discussed
in Paper III.

9. {\it 0226$-$1024.} This is the second-brightest high-z BAL QSO in our 
sample, with $V=16.9$. It has several distinct sub-components to its BALs.
Weak absorption features in the wavelength range 1740-1840\A may be due to
an Al III BAL. 0226$-$1024 was the target of higher resolution 
spectropolarimetric observations, discussed in Paper III. Preliminary
results were reported by \cite{o97}.

The continuum polarization rises to almost 2\% in the blue and the continuum 
PA is independent of wavelength. $P$ increases to 7\% in the C IV BAL and 
to 4\% in the Si IV BAL. $P$ does not rise in the sub-trough with lowest 
outflow velocity, suggesting that the corresponding BAL cloud covers the 
polarized light source more completely than the higher velocity BAL systems. 
The low-velocity associated absorber therefore has a different geometry and 
may have a different origin from the other BAL clouds. (This is discussed
in more detail in Paper III.) There are PA rotations of up to $-10\arcdeg$ in 
the BALs due to asymmetric partial coverage of the polarized continuum source. 
PA rotations in the opposite direction, in between the troughs, indicate the 
presence of resonantly scattered flux.

There is a large peak in polarized flux just redward of the N V BEL. It is
difficult to tell if this peak belongs to the polarized continuum or BEL 
because the continuum is not well defined at its wavelength. A similar peak is
seen in BAL QSO 1524+5147. The origin of these peaks could be Rayleigh 
scattering by neutral hydrogen (\cite{kf98}). This may also be
the source of the excess Ly$\alpha$+N V flux in the spectra of BAL QSOs 
relative to non-BAL QSOs (see, e.g.,~\cite{wmf91}). 

$P$ dips across all broad emission lines, including Fe II. There are deficits 
of polarized flux (below the continuum level) at the locations of the 
Ly$\alpha$, Si IV, and C IV BELs, indicating that the BELs are polarized at
a low level, nearly perpendicular to the continuum. In Paper III we suggest 
that this phenomenon is due to BEL photons resonantly scattered by the BAL 
region. 

10. {\it IRAS 07598+6508.} This is a low-redshift object with strong Fe II 
emission and a weak Na I BAL. There are few windows in the spectrum  
not heavily contaminated by broad emission lines. The continuum polarization 
and polarized flux rise to the blue.  We confirm the rise in $P$ in 
the Na I BAL seen by Hines \& Wills (1995). In addition, we find a PA 
rotation across the Na I BAL. This is similar to the high $P$ and PA 
rotations seen the C IV BAL of high-redshift BAL QSO. However, it is puzzling 
that such strong effects are seen in such a weak BAL. It is possible that
the true depth of the Na I BAL is filled in by an Fe II BEL. 

Our high S/N observations reveal that the BELs are polarized at a low level 
($\sim 0.5\%$) and at a PA different from the continuum. The PA rotates in
one direction in the blue wing and line core and in the opposite direction in
the red wing of H$\alpha$, an effect commonly seen in Seyfert 1s
(\cite{gm94}) and broad line radio galaxies (\cite{cot99}). The PA rotations 
are seen out to a velocity of $\pm 12,000$ km/s in the line wings. Most of the
line flux contributes to a positive rotation across the Fe II line blends. 

11. {\it 0842+3431 (CSO 203).} This BAL QSO has relatively narrow, detached 
BAL troughs. There is a weak Al III BAL, but it is contaminated by Fe II 
absorption from an intervening system at $z=1.161$. There are two more
intervening metal absorption systems. The continuum polarization is low 
($P_c=0.51\pm0.01$) and rises to the blue. $P$ rises to 3\% in the C IV BAL, 
a large contrast to the continuum polarization. The polarization drops at the 
N V emission line and there is no evidence for BELs in the polarized flux 
spectrum. 

The C IV trough depth increased by 7\% and the Si IV trough depth increased
by 16\% between the November 1994 and December 1995 observations. The Al III
BAL depth also increased noticeably. Neither the continuum nor peak C IV 
trough polarization changed significantly between the two epochs. This is
not surprising, since the changes in trough depth were small. Trough 
variability was also reported for this object by Barlow et al. (1992) over 
the years 1989-1991. It would be interesting to monitor this object with
sufficient precision to look for correlations among trough depth, trough
polarization, and continuum polarization. 

12. {\it 0856+1714.} The BALs are deep and show several sub-components, similar
to 0226$-$1024. One of these sub-components, showing up in both C IV and
Si IV, is unusual because it is redshifted with respect to the BEL
($\Delta v=+1700$ km/s). There is a weak Al III BAL. The polarimetric data 
are noisy since the object is faint and there were clouds. They are only 
useful for an estimate of the continuum polarization, which rises to the blue.

13. {\it 0903+1734.} The troughs of this BAL QSO are deep and show several 
sub-components. One of the narrow sub-components (at z=2.473) appears in
Al III absorption as well as in C IV and Si IV. Other than this, there is
no indication of an Al III BAL. There are two additional intervening metal 
absorption systems. $P$ rises in the C IV, Si IV, and Ly$\alpha$+N V BAL 
troughs, with a large contrast to the low continuum polarization. $P$ drops 
in the broad emission lines.  

14. {\it 0932+5006.} This object has multiple distinct BAL sub-troughs  
well separated in velocity. There are weak  Mg II and Al III BALs  
showing the same velocity structure as the C IV trough. However, the relative 
depths of the sub-troughs are different in the low ionization lines, 
suggesting either a velocity-dependent ionization or velocity-dependent 
line-of-sight covering fraction. 

$P$ rises in the multiple Si IV and C IV troughs, but the low-ionization 
troughs are too shallow to affect the polarization. The polarization rises
by only a factor of 2.3 in the deep portion of the C IV trough, a small 
contrast relative to other objects in our sample. The polarized continuum 
flux is reddened below 2000\A, similar to 0019+0107. The reddening 
is accompanied by a $15\arcdeg$ PA rotation across the continuum, another 
effect seen also in 0019+0107. The reddening and PA rotation may be due to 
patchy absorption of the scattered flux by dust in the scattering region.

15. {\it 1011+0906.} This is a low-ionization BAL QSO, with a prominent Al III
BAL and unusually strong Al III BEL. There are two intervening metal 
absorption line systems. The continuum polarization is a  moderate 2\%. The 
PA rotation at the blue end of the spectrum is of low significance. Both the
total flux and polarized flux spectra are redder than spectra of typical 
high-ionization BAL QSOs, suggesting dust extinction and reddening. There can 
only be a small difference between the reddening of the direct and scattered 
flux since there is no significant polarization wavelength dependence.

16. {\it 1212+1445.} The C IV BAL of this object has three narrow components 
at low velocity and a broad component at high velocity. The lowest 
velocity component is extremely narrow and unresolved in our spectrum, and
cuts into the peak of the C IV BEL. The deepest narrow component
($\Delta v=-2500$ km/s) is also visible in Al III and Mg II. However, the 
red component of the Al III doublet is confused with Fe II absorption from
an intervening system at $z=0.875$. The continuum polarization rises strongly 
to the blue. There is a $\sim10\arcdeg$ PA rotation in the red wing of
the C IV BEL at +6000 km/s, indicating polarized line emission and strongly 
suggesting a resonantly scattered component to the C IV flux. The other BELs 
are unpolarized and the polarization dilution signatures of the Fe II blends 
and C III] emission are prominent. 

The polarization changes strongly in magnitude and position angle across 
the broad portion of the C IV BAL trough. It appears that the narrow 
low-velocity components do not participate in this behavior. At $-$14,300 km/s
the polarization drops below the continuum level and the PA rotates
strongly ($120\arcdeg$). For this object, $Q^{\prime}$ is not a reliable 
estimate of P in the trough because of the large PA rotation. Therefore, we 
use $P=\sqrt{Q^2+U^2}$ (not debiased) in the vicinity of the trough, and 
$P=Q^{\prime}$ elsewhere. The net effect is that the high velocity BAL is 
deeper in polarized flux than in total flux and is also blue-shifted. 

17. {\it 1231+1320.} This is a low-ionization BAL QSO with a strong Al III BAL
and a large ratio of Al III to C III] emission. The spectrum is extremely 
unusual for a BAL QSO, for a number of reasons. There are multiple 
high-velocity sub-troughs in Al III, C IV and Si IV, with the highest velocity
sub-trough being the deepest. There is an overlap between the Si IV and C IV 
BALs because of their large velocity ranges. The Si IV BAL appears to be 
deeper in some of its sub-troughs than the C IV BAL. This would imply an 
unusually low ionization state for the corresponding BAL clouds. However, 
some of the extra depth in the Si IV trough may be due to extremely high 
velocity C IV absorption. The Ly$\alpha$+N V BEL is unusually weak and 
flat-topped. This is probably due to high-velocity Si IV BAL absorption. 
While the flux spectrum of 1231+1320 is quite interesting, the polarization is
very weak at $P_2=0.21 \pm 0.11$. Such a low polarization is only seen in two 
other BAL QSOs: 0043+0048 and 2201$-$1834. 

18. {\it 1232+1325.} This is a low-ionization BAL QSO like 1011+0906 and
1231+1320 with a strong Al III BAL and large Al III to C III] BEL flux ratio.
The C IV and Ly$\alpha$+N V BALs are saturated and go nearly black in the
low-velocity portion of the trough, similar to  0059$-$2735
(Fig. 4). The Si IV BAL shows a distinct sub-trough at
1290\A that is not apparent in the C IV BAL. The wavelength of this
structure does not coincide with the wavelengths of O I $\lambda$1302 or
C II $\lambda$1335 or their corresponding BALs, so it must be due to Si IV. 
Saturation of the C IV BAL probably smoothes out its profile. The 
high-velocity C IV BAL clouds are optically thick, but only partially cover 
the continuum sources. Unlike other low-ionization BAL QSOs in our sample, 
there is no indication of reddening. 

We identify this object as an HPQ, with continuum polarization 
$P_1=3.38 \pm 0.08\%$. This makes it a rare QSO in three senses--BAL,
low-ionization, and HPQ. There is little wavelength dependence of the 
continuum polarization, and there is no PA rotation. This, along with the 
lack of reddening in the total flux spectrum, is consistent with a simple 
electron scattering geometry. $P$ rises in the C IV, Si IV, and 
Ly$\alpha$+N V BALs, and perhaps in the Al III BAL. Consequently, the 
BALs are shallower in polarized flux than in total flux. $P$ rises distinctly
in the high-velocity Si IV sub-trough. $P$ drops strongly in all of the BELs, 
so they do not appear in the polarized flux spectrum. In all of its 
polarization properties, the low-ionization BAL QSO 1232+1325 is similar to 
the high-ionization HPQ BAL QSO in our sample. This is consistent with a 
similar geometry for the BAL outflow in low and high-ionization objects.  

19. {\it 1235+0857.} This object has deep, narrow BAL troughs cutting into the
broad emission lines. There are at least 7 additional weak  
sub-components extending to $-$24,100 km/s, giving the high velocity BAL
a ragged profile. There is a shallow Al III BAL at a velocity of 
$-$4100 km/s corresponding to the blue edge of the deep C IV sub-trough,
but it does not cut into the peak of the Al III BEL. As in 0932+5006, this 
suggests a velocity-dependent ionization. There is also a weak intervening 
metal absorber. 

The continuum polarization is moderately high ($P_c=2.53\pm0.07\%$) and 
constant with wavelength. This is the only object in our Keck sample (other
than 1331-0108, with low S/N) where the polarization  does not rise in
the troughs. However, there are large PA rotations in the blue
wings of the C IV, NV and Ly$\alpha$ BELs. These are significant at the 
3-4$\sigma$ level. It is difficult to tell if the rotations are due to 
polarized emission line flux or partial coverage of the polarized continuum 
source by the BAL. The rotation in the Ly$\beta$+O VI BAL is not significant
because it occurs where the trough goes black.

$P$ drops in the core and in the red and blue wings of the BELs, including the 
portions partially absorbed by the BALs. The BELs do not appear in the 
$Q'\times F$ spectrum, but there are narrow emission spikes in 
$U^{\prime}\times F$ corresponding to the PA rotations. The spike in 
$Q^{\prime}\times F$ at the wavelength of Ly$\beta$+O VI is due to noise. The 
BEL profiles are nicely recovered in the unpolarized (diluting) flux spectrum 
(Paper II). 

20. {\it 1235+1453.} This object has low polarization, with no significant
wavelength dependence. The deep part of the C IV BAL is lost in the dichroic 
split between the red and blue spectra.

21. {\it 1243+0121.} The ragged velocity profile of the C IV BAL is similar to
that of 0137$-$0153 (Fig. 3), with the deep portion of the 
trough only slightly detached from the BEL. There is a deep O VI+Ly$\beta$ 
BAL. The continuum polarization is moderate and wavelength-independent as well 
as can be judged from our noisy spectra. The polarization drops at all of 
the BELs, though this is of low significance in each instance.

22. {\it 1246$-$0542.} This is the brightest ($V=16.4$) high-redshift object 
in our sample. The BAL troughs consist of a shallow component which extends 
from $v=0-0.1c$ and a deep component centered at $-$16,000 km/s. Closer 
inspection reveals a number of narrower sub-troughs in the shallow region of 
the trough. There are weak Al III and Mg II BALs, centered at $-$16,000 km/s. 
There are also two intervening metal absorption line systems
($z=1.644$ and $z=1.200$). (Note that the narrow feature at 1270\A in the 
Si IV trough appears to be a combination of Al III and C IV absorption from 
the two intervening systems, and does not have a counterpart in the intrinsic
C IV trough.) 

The continuum polarization rises steadily to the blue and the polarized
continuum flux closely follows a power law with $\alpha=0.2$. The C IV and
Si IV broad emission lines have low polarization with PA perpendicular to 
the continuum PA, and are weakly present as dips in the polarized flux 
spectrum. We interpret their polarization as evidence for resonance 
scattering of BEL photons by the BAL clouds (Paper III). 

The polarization rises from 1.3\% in the continuum to 7.2\% in the C IV BAL, 
a large contrast. The BALs are shallow in polarized flux and appear to be 
blue-shifted. This indicates a very small line-of-sight covering fraction of 
the polarized light source by the BAL region. This covering fraction is also 
highly dependent on outflow velocity. The blue-shift could be a projection 
effect due to the different geometry of the direct and scattered rays. This 
object is discussed in detail in Paper III.
 
23. {\it 1331$-$0108.} This low-ionization BAL QSO was observed for a short 
time and has relatively low S/N. There is very strong Fe II emission, 
common to objects with strong low-ionization BALs. The deep Al III trough has 
a smaller velocity range than the C IV trough; it is missing the high-velocity
absorption tail. The Mg II trough is not as deep as the Al III trough and has 
an indistinct high velocity edge because it is partially filled in with Fe II 
emission. There is an intervening metal absorber. The continuum polarization 
is moderate, rises to the blue, and has a constant PA. Due to  the short 
exposure time there is little information about emission line and trough 
polarization.

24. {\it 1333+2840 (RS 23).} This classic object has P-Cygni line profiles. The
BELs have unusually large equivalent widths. Their peaks are also 
extremely narrow; the Si IV, Al III, and Mg II doublets are all resolved. 
We observed RS 23 on the suggestion of R. Goodrich, who hypothesized that 
large BEL equivalent widths may be an indicator of high polarization and 
attenuation of the continuum (\cite{g97}). This is indeed the most highly 
polarized object in our sample ($P_2=5.61\pm0.07\%$), and we identify it as a
new HPQ. 

The polarization rises by only 4\% in the C IV BAL, not nearly as 
dramatic a rise as in the polarization spectra of other BAL QSOs such 
as 0105$-$265. This is consistent with a continuum spectrum highly
attenuated and dominated by scattered light. Further absorption by the BAL 
has only a relatively small effect on the polarization. Dilution by the blue
wing of the C IV BEL and resonantly scattered photons from the BAL region may 
also reduce the trough polarization. There is a $+10\arcdeg$ PA rotation in 
the C IV BAL, peaking blueward of the deepest part of the trough and the
peak in $P$. As in 0226-1024, we attribute this rotation to uneven coverage
of the continuum scattering region by the BAL region.

Polarized C III] emission is marginally detected at the $2.1\sigma$ level, and
shows up as a dip in the polarized flux spectrum. There is a $-10\arcdeg$ PA 
rotation in the blue wing of the C IV BEL, in the opposite direction from the 
trough PA rotation. In addition, there is negatively polarized flux in the red
wing of C IV from photons resonantly scattered in the BAL region. The 
complicated polarization spectrum across the C IV line results from multiple 
sources of scattered light. We discuss this object in detail in Paper III.

25. {\it 1413+1143 (Cloverleaf).} This BAL QSO is lensed into a quadruple 
source by an intervening galaxy and perhaps a distant cluster of galaxies
(\cite{mss88}; \cite{kam98}). Magnification by the lens is proving to be a 
useful tool for studying the structure of the QSO. The line profiles are 
P-Cygni like, with more velocity sub-structure than RS 23 (see above), and the 
C IV trough is nearly black. There is a weak Al III BAL at a velocity 
corresponding to the deepest part of the C IV BAL. There are 3 intervening 
metal-line absorption systems ($z=$1.354, 1.437, and 1.659) apparent in our 
spectrum. 

The polarization properties of 1413+1143 have been reported by Goodrich \&
Miller (1995), including evidence of variable continuum polarization. In spite
of the variability, they suggest a scattering origin for the polarization. The 
continuum polarization rises to the blue and the continuum PA is constant
with wavelength. The polarization rises strongly in the BAL troughs, 
including Ly$\alpha$, N V, Si IV, C IV, and perhaps P V. $P$ reaches the very
high value of $17\pm 5\%$ in the C IV trough. This is one of the first BAL 
QSOs to show $P$ rising in the absorption troughs (\cite{gm95}). There is also
a consistent PA rotation of about $-20 \arcdeg$ across the deep BALs. 

The permitted BELs are polarized (at a lower level than the continuum), 
while the semi-forbidden C III] line is unpolarized. This is opposite to the 
polarization behavior in 2225$-$0534 and 0019+0107, where only the 
C III] BEL is polarized. The permitted BELs in 1413+1143 are probably 
polarized by scattering in the same region as the continuum since they have
a similar PA, and the scatterers must be relatively cold since they preserve
the BEL width.

26. {\it 1524+5147 (CSO 755).} This is a very bright BAL QSO with multiple, 
shallow troughs. There is Mg II absorption from 2 intervening systems 
($z=$1.395, 1.453). The polarization properties of CSO 755 were first reported
by Glenn et al. (1994). The continuum polarization is high (3.5\%) and rises 
to the blue. The BAL and BEL profiles are quite different from those of the 
other HPQ in our sample (RS 23, 1232+1325, and PHL 5200). The BALs of CSO 755 
are detached, narrow, and shallow, while the BALs of the other HPQ  have deep 
P-Cygni profiles. The BELs have normal equivalent widths, in contrast to the 
high equivalent width BELs of the other HPQ. As we discuss further in Paper 
II, there is no correlation between BAL and BEL properties and continuum 
polarization in our sample as a whole. 

The polarization rises by only 1-2\% in the Ly$\alpha$ and C IV troughs, 
because they are not very deep. There are no polarization increases in the two 
lowest-velocity associated absorption systems, indicating that they cover the 
polarized light source more completely than the other absorption systems.
However, the low-velocity systems only partially cover the BEL region 
(see Paper III).

There is a $-5\arcdeg$ PA swing across the Ly$\alpha$+N V BAL and a 
$+5\arcdeg$ rotation across the corresponding BEL. This effect is absent in
the C IV BAL and BEL. In addition, the polarized flux peaks in the red wing of
NV. As in 0226$-$1024, this may be due to polarization by Rayleigh scattering 
in the Ly$\alpha$ BEL. There is a small deficit of polarized flux in the red 
wings of all permitted lines because they are polarized at a low level, 
perpendicular to the continuum.

27. {\it 1700+5153 (PG, IRAS).} This low-redshift BAL QSO has an Mg II BAL and
other BALs in the observed UV (e.g.,~\cite{pb85}). There is no apparent Na
I BAL. PG 1700+5153  belongs to a class of AGN with strong Fe II emission and 
extremely weak [O III] emission, including 07598+6508, Mrk 231, and 
14026+4341 (\cite{bm92}). This is one of the few known low-redshift BAL QSOs, 
difficult to identify because the BALs are in the UV. It is bright 
and nearby, allowing high S/N spectropolarimetry of it from Palomar. The 
optical continuum polarization is low (1.1\%), and increases mildly to the 
blue. The same polarization behavior is seen in the UV spectra 
of high-z BAL QSOs, making it an extremely broad-band phenomenon (900-6400\A).
$P$ drops in H$\beta$ and the Fe II emission lines, so they do not show up in 
the polarized flux spectrum. 

There is a moderate PA rotation ($12\arcdeg$) across the continuum. This
may be partly due to interstellar polarization, as large as 0.3\% at the 
Galactic latitude of 1700+5153 ($b=37\fdg 8$). Depending on the orientation of
the interstellar polarization, it could cause a rotation of up to $7\arcdeg$. 
It is necessary to measure a nearby interstellar polarization probe star to 
determine the proper correction. Schmidt \& Hines (1999) also measure a 
$10\arcdeg$ PA rotation in the continuum, confirming our result, and find a 
PA rotation across H$\alpha$, outside of our spectral coverage. 

This is one of the very few BAL QSOs bright and near enough to have a resolved
radio structure. It consists of two unresolved sources separated by 
$1\arcsec$ at $PA\sim 100\arcdeg$ (\cite{kss94}). This can 
neither be considered parallel nor perpendicular to the optical polarization 
vector ($PA=53\arcdeg$). This is contrary to the suggestion of Goodrich \& Miller 
(1995) that the polarization PA and radio axis are parallel, as would be
expected for scattering in the equatorial plane. If the radio structure is 
due to a weak double-lobed jet, it may be deflected and may not give a 
reliable indication of the spin axis of the central black hole. The 
misalignment of the radio jet and polarization axis would not surprising 
in this context.

Ground-based and HST images have shown an interacting companion galaxy to the 
QSO (\cite{scc98}, and \cite{hlt99}), located $2\arcsec$ to the north 
at $PA\simeq 0\arcdeg$ (roughly perpendicular to the radio axis). Keck 
spectroscopy by Canalizo \& Stockton (1997) shows a strong starburst component
in the companion that may be indirectly related to the QSO activity, with a 
flux $\sim 1000\times$ weaker than the QSO. We do not see this stellar 
component in our spectrum, consistent with its faintness and location on the 
edge of our slit. 

28. {\it 2154$-$2005.} This object has very shallow BALs that appear to have 
weakened since the observations of Weymann et al. (1991). This is similar to 
the case of 0025$-$0151, which also has weak variable BALs. The continuum 
polarization of 2154$-$2005 is low (0.9\%) and rises gradually to the blue, 
while the PA is independent of wavelength. This object has typical BAL QSO 
polarization characteristics in spite of its unusually weak BALs. It may be 
considered a transition object between BAL QSOs and non-BAL QSOs. 

29. {\it 2201$-$1834.} (Fig. 5). This is an unusual BAL QSO 
in a few respects. The spectrum is highly reddened. The C IV BAL has 
a very low equivalent width (16.3\A) and there is no Mg II BAL, 
atypical of reddened BAL QSOs. Finally, the continuum has very low 
polarization ($P=0.16 \pm 0.07\%$), unexpected for an object with 
high extinction. Perhaps we are viewing a high-ionization BAL QSO 
through a foreground dust screen in the host galaxy ISM that is not directly 
connected to the AGN phenomenon. The dust grains are not aligned, or they 
would induce a measurable polarization.

30. {\it 2225$-$0534 (PHL 5200).} Spectropolarimetric measurements of this 
object were reported by Cohen et al. (1995), Goodrich \& Miller (1995), and 
Stockman, Angel, and Hier (1981). While often called the prototypical BAL QSO,
it is not well representative of its class. Its BELs have extreme equivalent 
widths like 1333+2840 (see above). This may indicate an attenuated continuum 
seen primarily in scattered light. Its deep, broad BALs start at 0 km/s and 
absorb the blue wings of the BELs. There is a modest Mg II BAL corresponding 
to the deepest portion of the saturated C IV trough. 

PHL 5200 is the second-most highly polarized BAL QSO in our sample, with 
$P$ rising to 5\% in the blue. This is another extreme characteristic for its 
class. The electric vector PA is roughly constant with wavelength. The 
polarization increases to 10\% in the Si IV and C IV BALs, indicating an 
excess of scattered light unabsorbed by the BAL clouds. This is one of the 
first objects where this phenomenon was detected (\cite{cot95}). 

$P$ drops in the emission lines, including the blended Fe II features, 
but dilution by Fe II is insufficient to completely account for the wavelength 
dependence of the continuum polarization. There is residual polarized flux 
and a small PA rotation at the C III] BEL, but the C IV BEL is unpolarized. 
The difference in polarization between C III] and C IV has been attributed to 
resonance scattering in the low-optical depth C III] line (\cite{cot95}).

\acknowledgments

This paper is derived in large part from the Ph.D. thesis of P.~M. Ogle,
conducted at the California Institute of Technology. We thank Bev Oke and 
Judy Cohen for all of their hard work in building LRIS and assistance with 
its polarimetry mode. We are grateful to Angela Putney, John Gizis, Bradford 
Behr, Greg Harper, and Patrick Shopbell for assisting with the Palomar 
observations. We thank Nahum Arav, Tom Barlow, Roger Blandford, Ray Weymann,
and the referee, Bev Wills, for interesting discussions which improved this 
paper. The W. M. Keck Observatory is operated as a scientific partnership 
between the California Institute of Technology, the University of California, 
and the National Aeronautics and Space Administration. It was made possible by
the generous financial support of the W. M. Keck Foundation. This research has 
made use of the NASA/IPAC Extragalactic Database, which is operated by the Jet
Propulsion Laboratory, Caltech under contract with NASA. Research by HDT at 
LLNL is supported by the DOE under contract W7405-ENG-48. Work on this paper
was supported, in part, by NSF grants AST-9121889 and AST-9117100.

\appendix
\section{Palomar Instrumental Polarization}\label{sec:inspol}

There is a large instrumental polarization correction for the Palomar 
observations taken after 10/94. Figure 6 shows the mean 
instrumental polarization curves for 11/95-10/96, derived from observations 
of null polarization standards. The curves are displayed in instrumental 
coordinates for a slit position angle of $PA=0\arcdeg$. The blue curve is 
corrected for an extra reflection at the dichroic filter. All BAL QSO data 
were corrected by subtracting a spline fit to the raw instrumental 
polarization curve from the Stokes parameters. The instrumental polarization 
correction is of the same magnitude as the typical BAL QSO continuum 
polarization, so it is important to use an accurate curve. The instrumental 
polarization rises from 0.6\% at 8100\A to 2\% at 3700\A. Except for some 
noise just blueward of the dichroic split, the curves match across the 
dichroic split. 

The instrumental polarization varied less than 0.2\% from run to run, and the
small variations were probably due to weak polarization in the null standards.
The one exception is the 10/94 observing run, where there was no measurable
instrumental polarization. Typically 2 or 3 null standards were averaged to 
obtain the correction curve for each run, reducing the effect of slight
interstellar polarization in some standards. It is assumed that the correction
is additive and does not depend on the true polarization of the object. This
is similar to interstellar polarization correction, where the intervening
dust can be approximated by a single partial-polarizing screen. The 
second-order corrections are less than the linear correction by a factor of
the object polarization $P_c$ (\cite{g86}), so are of order 0.01\% in our 
observations. We checked the accuracy of the correction by comparing 
Palomar and Keck observations of the same objects. These agree well, giving 
confidence in the correction (see Fig. 2). 

The Double Spectrograph and polarimeter are mounted on a ring at Cassegrain 
focus. When the ring is rotated, the PA of the instrumental polarization 
rotates by the same amount. This means the source of instrumental polarization
is not the spectrograph or polarimeter, but either the primary or secondary 
mirror of the telescope. The 10/94 observing run showed little or no 
instrumental polarization. It is possible that cleaning and subsequent 
realignment of the telescope secondary mirror is to blame for the large 
increase in instrumental polarization between 10/94 and 11/95. If the 
secondary mirror is not aligned perfectly normal to the primary mirror, then 
the projection of the secondary will be slightly elliptical and induce a net 
polarization.  

The chromatic nature of the instrumental polarization suggests that it is not 
simply due to mirror misalignment. Another possibility is uneven 
re-aluminization of the primary mirror, which can induce a net polarization. 
In this case the wavelength dependence could come from tarnished spots which 
selectively absorb blue light. In fact, re-aluminization of the 200 inch 
primary at Palomar is a very tricky business, and sometimes gives less than 
uniform coverage. There are similar difficulties with instrumental 
polarization in Lick Observatory spectropolarimetry (\cite{m96}), which is 
also due to irregular aluminization of the telescope mirrors. The wavelength 
dependence of the instrumental correction is somewhat different at Lick; it 
does not increase monotonically to the blue as at Palomar. Also, the magnitude 
of the effect is somewhat less ($\sim$ 1.5\%) at Lick. 

\section{Interstellar Polarization}

A prevalent concern in polarimetry is contamination of the Stokes parameters
by intervening dust. Polarization can be induced by selective
absorption by magnetically aligned dust grains. Interstellar dust 
polarization in the Galaxy should have only a small effect on the 
Stokes parameters since all objects have a Galactic latitude of $b>30\arcdeg$.
Table 9 gives the Galactic latitude for each object in 
the Keck and Palomar samples. The maximum interstellar polarization $P_{max}$
was estimated using the prescription of Serkowski et al. (1975) and the 
extinction values of Burstein \& Heiles (1984), which are listed in NED. 
\footnote{NED is the NASA Extragalactic Database, operated by IPAC.} We used
the formula $P_{max}=9E(B-V)=(9/4)A_B$ to convert from extinction values to 
polarization. The maximum expected interstellar polarization is typically less
than 0.2\%.

We also list the polarization $P_s$, position angle $PA_s$, distance $d$,
and separation $d\theta$ of the closest field star within $5\arcdeg$ of each 
QSO from ~\cite{mf70}. This is at best a crude estimate of the interstellar 
polarization in the direction of the QSO. Some of these stars are too close to
the observer to sample the total dust column in the Galactic disk. Ideally, we 
would like to take spectropolarimetry of several stars within a few arc minutes
of the QSO which have large spectroscopic distances ($>0.5$kpc). Then it would
be possible to make a more reliable correction for the interstellar 
polarization (See, e.g., ~\cite{t95}). The number of observations required for
this would be prohibitively large and impractical for the current study. We 
consider any measured QSO polarization greater than 0.2\% to be intrinsic to 
the QSO. However, we also investigate the possibility of interstellar 
contamination in objects which show rotation of the electric vector PA with 
wavelength (Paper II).

Another possible source of dust polarization is the ISM of intervening  
galaxies. Many of the BAL QSOs in our sample show intervening metal absorption 
line systems. Typically, the impact parameter of these systems is large,
so we don't expect a large amount of extinction and associated dust 
polarization. QSOs with damped Ly$\alpha$ absorbers at small impact parameters
typically have an extinction at 1500\A of $A<0.1$ magnitudes (\cite{pks97}). 
We expect even less extinction and negligible interstellar polarization
from the lower column density metal absorption line systems. However, 
1413+1143 is lensed by a foreground galaxy, so dust absorption by the lensing 
galaxy may be a consideration. Dust absorption by the QSO host galaxy may be
important in some QSOs with reddened spectra, and this will be considered
along with other intrinsic polarization sources in Paper II.

\clearpage

\figcaption[Keck PA calibration curve]{Keck PA calibration curve, showing 
rotation of half-wave plate fast axis with wavelength.}

\figcaption[Comparison of Palomar and Keck polarization measurements.]
{Comparison of Palomar and Keck polarization measurements. Note that there 
are no systematic differences between the two data sets. Also, there is no 
evidence of intrinsic variability between the sets.}

\figcaption{Spectropolarimetry of 29 BAL QSOs (3a-3o). From top to bottom the 
four panels display (1) total flux $F_{\lambda}$, (2) polarization 
fraction (rotated Stokes parameter $Q^{\prime}$), (3) polarized flux 
$Q^{\prime}\times F$, and  (4) position angle $\theta$. Total and polarized 
fluxes are plotted in units of $10^{-16}$ erg s$^-1$ cm$^{-2}$ \A$^{-1}$. 
Prominent emission lines are indicated in the top panel by tick marks. The 
horizontal line in panel 4 is the mean PA listed in Table 5. Since 0043+0048 
(UM 275) and 1231+1320 are unpolarized, we display $F_{\lambda}$ and the 
unrotated Stokes $Q$ and $U$ parameters only. Individual objects are discussed
in the text.}

\figcaption[Palomar flux spectra, objects repeated at Keck]{Palomar 
spectra of objects re-observed at Keck. Flux units are 
$10^{-16}$erg s$^{-1}$ cm$^{-2}$ \A$^{-1}$. The Palomar spectra extend 
further to the blue than the Keck spectra.}

\figcaption[Flux spectra, polarization non-detections]{ 
Flux spectra of polarization non-detections. Flux units are 
$10^{-16}$erg s$^{-1}$ cm$^{-2}$ \A$^{-1}$. Note that 
spectra of 0043+0048 and 1231+1320 are presented in Fig. 3.}

\figcaption[Palomar instrumental polarization]{Mean Palomar instrumental
   polarization correction fraction for Stokes Q and U parameters. Note that 
   these are in instrumental coordinates, not sky coordinates.}

\clearpage

\begin{deluxetable}{llcllrcl}
 \tablewidth{0pt}
 \tablecaption{BAL QSO observation log, Keck}
 

\tablehead{\colhead{IAU (B1950)} & \colhead{name} & \colhead{z\tablenotemark{a}} &
           \colhead{V\tablenotemark{b}} & \colhead{UT date} & 
           \colhead{t\tablenotemark{c}} & \colhead{PA\tablenotemark{d}} & 
           \colhead{notes}}

\startdata
0019$+$0107  &UM 232  & 2.123 & 17.6 & 1995 07 29 & 6240 & 315   &   \nl    
             &        &       &      & 1995 12 15 & 6000 &  35.4 &   \nl  
0043$+$0048  &UM 275  & 2.146 & 18.4 & 1994 10 29 &10800 & 305   &\tablenotemark{np} \nl 
0059$-$2735  &\nodata & 1.593 & 17.0 & 1994 12 30 & 6240 &  20   &   \nl  
             &        &       &      & 1995 07 28 & 6240 &   0   &   \nl  
0105$-$265\tablenotemark{e}   &\nodata & 3.488 & 17.3 & 1994 08 03 & 3600 &   0   & \nl  
             &        &       &      & 1994 10 28 & 3600 &   0   &   \nl 
             &        &       &      & 1994 11 06 & 3600 &  33.4 &\tablenotemark{np} \nl
0137$-$0153  &UM 356  & 2.234 & 17.8 & 1995 12 17 & 6000 & 227   &\tablenotemark{np} \nl   
0146$+$0142  &UM 141  & 2.892 & 17.8 & 1994 11 06 & 3600 &  60.6 &\tablenotemark{np} \nl
             &        &       &      & 1995 12 15 & 4800 &  50   &   \nl 
0226$-$1024  &\nodata & 2.256 & 16.9 & 1994 12 31 & 6000 &  45   &   \nl  
             &        &       &      & 1996 10 05 & 7200 & 181.5 &   \nl 
             &        &       &      & 1996 10 06 & 7200 & 181.5 &\tablenotemark{np} \nl 
07598$+$6508\tablenotemark{e} &IRAS    & 0.148 & 14.5 & 1995 01 28 & 2400 &  45   & \tablebreak
0842$+$3431\tablenotemark{e}  &CSO 203 & 2.130 & 17.2 & 1994 11 06 & 3600 & 270   &\tablenotemark{np}  \nl 
             &        &       &      & 1995 12 17 & 6000 & 160   &\tablenotemark{np} \nl
0856$+$1714\tablenotemark{e}  &\nodata & 2.327 & 19.0 & 1994 11 07 & 3600 & 283   &\tablenotemark{np}  \nl 
0903$+$1734  &\nodata & 2.773 & 17.7 & 1996 04 17 & 4800 &  90   &   \nl  
0932$+$5006  &\nodata & 1.911 & 17.2 & 1995 12 15 & 6000 & 215   &   \nl
1212$+$1445  &\nodata & 1.626 & 17.4 & 1996 05 19 & 7200 &   0   &\tablenotemark{np} \nl
1232$+$1325  &\nodata & 2.361 & 18.1 & 1998 05 26 & 3600 & 325   &\tablenotemark{np} \nl
1235$+$0857  &\nodata & 2.887 & 17.7 & 1996 05 20 & 3600 &  50   &\tablenotemark{np} \nl
1246$-$0542  &\nodata & 2.226 & 16.4 & 1994 12 30 & 7050 & 315   &   \nl
1331$-$0108  &UM 587  & 1.874 & 17.4 & 1996 05 19 & 1680 &  57   &\tablenotemark{np} \nl
1333$+$2840  &RS 23   & 1.910 & 18.7 & 1995 01 28 & 4800 & 270   &   \nl   
             &        &       &      & 1996 04 17 & 6000 & 100   &   \nl 
1413$+$1143  &Cloverleaf & 2.545 & 16.8 & 1996 05 19 & 3600 &  68   &\tablenotemark{np} \nl   
1524$+$5147\tablenotemark{e}  &CSO 755 & 2.88  & 17.1 & 1996 04 17 & 3600 & 132   & \nl
2225$-$0534  &PHL 5200& 1.980 & 18.4 & 1994 08 03 & 3600 & 180   &   \nl
             &        &       &      & 1994 10 28 & 3600 &  35   &  \nl
\enddata

\tablenotetext{a}{redshift (\cite{wmf91})}
\tablenotetext{b}{apparent V-band magnitude (\cite{b93})}
\tablenotetext{c}{exposure time (sec)}
\tablenotetext{d}{spectrograph slit PA}
\tablenotetext{e}{Object not in the Weymann et al. (1991) sample}

\tablenotetext{np}{Non-photometric (clouds or fog)}

\end{deluxetable}
\clearpage

\begin{deluxetable}{llcllrcl}
 \tablewidth{0pt}
 \tablecaption{BAL QSO observation log, Palomar}
 

\tablehead{\colhead{IAU (B1950)} & \colhead{name} & 
           \colhead{z\tablenotemark{a}} &
           \colhead{V\tablenotemark{b}} & \colhead{UT date} & 
           \colhead{t\tablenotemark{c}} & \colhead{PA\tablenotemark{d}} & 
           \colhead{notes}}

\startdata
0021$-$0213 & \nodata  & 2.293 & 18.7 & 1996 10 17 & 7200 &   0 & \tablenotemark{np} \nl
0025$-$0151 & UM 245   & 2.075 & 18.1 & 1996 10 18 & 3600 &  25 & \tablenotemark{np} \nl
0043$+$0048 & UM 275   & 2.146 & 18.4 & 1994 10 27 & 3600 &  25 &    \nl
0059$-$2735 & \nodata  & 1.593 & 17.0 & 1994 10 27 & 5400 &   0 &    \nl
0119$+$0310\tablenotemark{e} & NGC470D8 & 2.10  & 18.1 & 1996 10 18 & 3600 &  30 & \tablenotemark{np} \nl
0137$-$0153 & UM 356   & 2.234 & 17.8 & 1995 11 24 & 6480 &  30 & \tablenotemark{np} \nl
0145$+$0416 & UM 139   & 2.028 & 18.3 & 1996 10 17 & 6000 &  20 & \tablenotemark{np} \nl
0226$-$1024 & \nodata  & 2.256 & 16.9 & 1994 10 27 & 3600 &  10 &    \nl
0846$+$1540 \tablenotemark{e} & \nodata  & 2.912 & 18.0 & 1996 10 18 & 3360 & 305 & \tablenotemark{np} \nl
0903$+$1734 & \nodata  & 2.773 & 17.7 & 1995 11 24 & 6000 & 310 & \tablenotemark{np} \nl
0932$+$5006 & \nodata  & 1.911 & 17.2 & 1995 11 23 & 7500 & 220 & \tablenotemark{np} \nl
1011$+$0906 & \nodata  & 2.262 & 17.7 & 1996 02 23 & 7200 &  10 & \tablenotemark{np} \nl
1212$+$1445 & \nodata  & 1.626 & 17.4 & 1996 05 12 & 3600 &  20 &    \nl
1231$+$1320 & \nodata  & 2.383 & 18.0 & 1996 05 12 & 6000 &  45 &    \nl
1232$+$1325 & \nodata  & 2.361 & 18.1 & 1996 02 23 & 6000 &   0 & \tablenotemark{np} \nl
1235$+$0857 & \nodata  & 2.887 & 17.7 & 1996 05 13 & 3600 &  34 &    \nl
1235$+$1453 & \nodata  & 2.686 & 18.4 & 1996 05 13 & 7800 &  30 &    \nl
1243$+$0121 & \nodata  & 2.790 & 18.5 & 1996 02 23 & 5400 &  30 & \tablenotemark{np} \nl  
1442$-$0011 & \nodata  & 2.216 & 18.2 & 1996 05 12 & 6000 &  35 &    \nl
1443$+$0141 & \nodata  & 2.444 & 18.2 & 1996 05 13 & 7200 &  35 &    \nl
1700$+$5153\tablenotemark{e} & PG, IRAS & 0.292 & 15.1 & 1996 05 13 & 2400 & 130 &  \nl
2154$-$2005 & \nodata  & 2.029 & 18.1 & 1994 10 27 & 3600 &   0 &    \nl
2201$-$1834 & \nodata  & 1.814 & 17.6 & 1995 11 25 & 5400 &  20 & \tablenotemark{np} \nl
            &          &       &      & 1996 10 17 & 5400 &   5 & \tablenotemark{np} \nl
2350$-$0045 & \nodata  & 1.624 & 18.6 & 1996 10 18 & 7200 &  10 & \tablenotemark{np} \nl
\enddata
\tablenotetext{a}{redshift (\cite{wmf91})}
\tablenotetext{b}{apparent V-band magnitude (\cite{b93})}
\tablenotetext{c}{exposure time (sec)}
\tablenotetext{d}{spectrograph slit PA}
\tablenotetext{e}{Object not in the Weymann et al. (1991) sample}

\tablenotetext{np}{Non-photometric (clouds or fog)}

\end{deluxetable}

\begin{deluxetable}{cccccc}
\tablecaption{Polarized standard stars, Keck observations}
\tablehead{\colhead{Star} & \colhead{Band} & \colhead{$P\%$} & 
           \colhead{$\sigma$} & \colhead{PA} & \colhead{$\sigma$}}
\startdata
HD 155528   &  V &   5.00 & 0.17  &  92.61 & 0.37  \nl
HD 245310   &  B &   4.38 & 0.004 & 145.97 & 0.40  \nl
VI Cyg 12   &  V &   9.16 & 0.06  & 116.41 & 0.13  \nl
Hiltner 102 &  V &   5.18 & 0.10  &  73.20 & 0.80  \nl
HD 251204   &  V &   4.98 & 0.05  & 151.60 & \nodata \nl
\enddata


\tablecomments{No PA uncertainty is listed for HD 251204, because it was 
 observed on only one occasion.}

\end{deluxetable}
  
\begin{deluxetable}{lccc}
\tablecaption{Null polarization standard stars, Keck observations} 


\tablehead{\colhead{Star} & \colhead{Date} & \colhead{$Q\%$} & \colhead{$U\%$}}
\startdata
G191B2B     & 1994 12 &   0.052 &   0.009 \nl
            & 1995 12 &$-$0.050 &$-$0.045 \nl
            & 1996 10 &   0.002 &   0.037 \nl
BD +28 4211 & 1995  7 &$-$0.069 &$-$0.079 \nl
            & 1996 10 &   0.033 &$-$0.111 \nl
            & 1998  5 &$-$0.055 &$-$0.043 \nl
BD +32 3739 & 1995  7 &   0.057 &$-$0.096 \nl
\hline
            & mean     &  $-$0.004 &$-$0.047 \nl
            & $\sigma$ &   0.05  &   0.05  \nl
\enddata
\tablecomments{Stokes $Q$ and $U$ parameters were measured in the
              4000-7000\A band.}
\end{deluxetable}

\begin{deluxetable}{llcccc}
\tablecaption{BAL QSO continuum polarization, Keck measurements}

\tablehead{\colhead{IAU (B1950)} & \colhead{Name} & \colhead{$P_1\%~\pm$} & 
           \colhead{$P_2\%~\pm$} & \colhead{$P_c\%~\pm$} & 
           \colhead{PA$~\pm$}}
\startdata
 0019$+$0107  &UM 232  &0.94 0.03 & 1.10 0.03 &0.98 0.02 & 35.0 0.5 \tablenotemark{r} \nl
 0043$+$0048  &UM 275  &0.05 0.04 \tablenotemark{a}& 0.11 0.04 \tablenotemark{a}&0.11 0.06 \tablenotemark{a}& \nodata \tablenotemark{a}   \nl
 0059$-$2735  &\nodata &1.58 0.10 & 1.62 0.06 &1.49 0.02 &171.2 0.3  \nl
 0105$-$265   &\nodata &1.45 0.05 & \nodata \tablenotemark{b}  &2.41 0.08 &134.8 1.2  \nl
 0137$-$0153  &UM 356  &1.06 0.07 & 0.81 0.08 &1.09 0.05 & 56.0 1.3  \nl
 0146$+$0142  &UM 141  &1.12 0.03 & 1.15 0.08 &1.24 0.02 &133.1 0.5  \nl
 0226$-$1024  &\nodata &1.80 0.02 & 1.73 0.02 &1.81 0.01 &167.1 0.2  \nl
 07598$+$6508 & IRAS   & \nodata \tablenotemark{b} & \nodata \tablenotemark{b}  &1.709 0.005&123.05 0.08 \nl
 0842$+$3431  &CSO 203 &0.52 0.02 & 0.55 0.03 &0.51 0.01 & 27.1 0.6  \nl
 0856$+$1714  &\nodata &0.88 0.13 & 0.91 0.10 &1.02 0.10 &165.7 2.8  \nl
 0903$+$1734  &\nodata &0.52 0.04 & 0.44 0.05 &0.67 0.02 & 62.9 0.6  \nl
 0932$+$5006  &\nodata &1.20 0.02 & 1.19 0.03 &1.11 0.02 &168.7 0.5 \tablenotemark{r} \nl
 1212$+$1445  &\nodata &2.05 0.06 & 1.88 0.06 &1.49 0.03 & 17.3 0.6  \nl
 1232$+$1325  &\nodata &3.38 0.08 & 3.30 0.06 &3.19 0.04 & 93.9 0.3  \nl
 1235$+$0857  &\nodata &2.76 0.06 & 2.65 0.09 &2.53 0.07 & 25.1 0.7  \nl
 1246$-$0542  &\nodata &1.31 0.02 & 1.11 0.02 &1.26 0.01 &132.4 0.1  \nl
 1331$-$0108  &UM 587  &1.76 0.13 & 1.55 0.09 &1.56 0.14 & 33.8 2.2  \nl
 1333$+$2840  &RS 23   &5.41 0.11 & 5.61 0.07 &4.67 0.02 &161.5 0.1  \nl
 1413$+$1143  &Cloverleaf &1.49 0.04 & 1.24 0.08 &1.52 0.04 & 55.7 0.9 \nl
 1524$+$5147  &CSO 755 &3.56 0.03 & 3.42 0.05 &3.49 0.01 & 98.6 0.1  \nl
 2225$-$0534  &PHL 5200&4.98 0.05 & 4.63 0.06 &4.26 0.02 &162.0 0.2  \nl
\enddata
\tablenotetext{a}{UM 275 has insignificant polarization, so we list 
   $P=\sqrt{Q^2+U^2}$ instead of $P=Q^{\prime}$; and the PA is highly 
   uncertain, so we do not include it.}
\tablenotetext{b}{There is no measurement of $P_2$ for 0105$-$165 because of 
   its high redshift, and no measurement of $P_1$ or $P_2$ for 07598+6508 
   because of its low redshift.}
\tablenotetext{r}{Significant continuum PA rotation}
\tablecomments{$P_1$ and $P_2$ are measured in two narrow wavelength bands 
   (1600-1840\A and 1960-2200\A, rest) to either side of C III] $\lambda$1909.
   $P_c$ and PA are the broad-band 'white light' polarization and PA 
   (4000-8600\A, observed). Uncertainties are listed as standard deviations of
   the mean within the given band; they are are slightly larger than the 
   formal uncertainty from photon statistics, due to polarization variation 
   within the wavelength bands.  }

\end{deluxetable}

\begin{deluxetable}{llccc}
\tablecaption{BAL QSO continuum polarization, Palomar measurements}
\tablehead{\colhead{IAU (B1950)} & \colhead{Name} & \colhead{$P_1$\% $\pm$} & 
           \colhead{$P_2$\% $\pm$} & \colhead{PA $\pm$}}

\startdata
0025$-$0151 & UM 245   &  \nodata  & 0.57 0.16 &  108  9 \nl
0059$-$2735 & \nodata  & 1.45 0.17 &  \nodata  &  168  8 \nl
0137$-$0153 & UM 356   &  \nodata  & 0.46 0.13 &   71 15 \nl
0145$+$0416 & UM 139   &  \nodata  & 2.14 0.10 &  126  1\tablenotemark{r} \nl
0226$-$1024 & \nodata  &  \nodata  & 1.84 0.10 &  166  1 \nl
0903$+$1734 & \nodata  & 0.53 0.06 &  \nodata  &   62  6 \nl
0932$+$5006 & \nodata  & 1.17 0.30 & 1.25 0.07 &  172  1\tablenotemark{r} \nl
1011$+$0906 & \nodata  &  \nodata  & 2.00 0.08 &  132  2 \nl
1212$+$1445 & \nodata  & 2.17 0.34 &  \nodata  &   24  1 \nl
1232$+$1325 & \nodata  &  \nodata  & 3.49 0.19 &   92  3 \nl
1235$+$0857 & \nodata  & 2.16 0.15 &  \nodata  &   21  2 \nl
1235$+$1453 & \nodata  & 0.75 0.13 &  \nodata  &  175 12 \nl
1243$+$0121 & \nodata  & 1.68 0.20 &  \nodata  &  136  4 \nl
1700$+$5153 & PG, IRAS &  \nodata  & 1.07 0.04\tablenotemark{a}&   53  1\tablenotemark{r} \nl  
2154$-$2005 & \nodata  &  \nodata  & 0.90 0.17 &  142  8 \nl
\enddata

\tablenotetext{a}{1700+5153 was measured in the band 3700-4100\A, rest.}
\tablenotetext{r}{Significant continuum PA rotation}
\tablecomments{$P_1$ and $P_2$ are the polarizations in two narrow wavelength 
bands (1600-1840\A and 1960-2260\A, rest) to either side of C III] 
$\lambda$1909. For most objects, only one of the two bands is available, due 
to the dichroic split. PA is the corresponding electric vector position 
angle in degrees (except for 0932+5006, where PA is the average of the two 
bands).}

\end{deluxetable}

\begin{deluxetable}{llccc}
\tablecaption{BAL QSO continuum polarization, Palomar nondetections}
\tablehead{\colhead{IAU (B1950)} & \colhead{Name} & \colhead{$Q$\% $\pm$} & 
           \colhead{$U$\% $\pm$} & \colhead{band}\tablenotemark{a}}
\startdata
0021$-$0213 & \nodata  &   0.28 0.12 &$-$0.42 0.23 &  2   \nl
0043$+$0048 & UM 275   &   0.12 0.19 &   0.06 0.19 &  2   \nl
0119$+$0310 & NGC470D8 &   0.34 0.16 &   0.00 0.20 &  2   \nl
0846$+$1540 & \nodata  &   0.30 0.33 &   0.01 0.36 &  1   \nl
1231$+$1320 & \nodata  &$-$0.16 0.07 &$-$0.14 0.10 &  2   \nl 
1442$-$0011 & \nodata  &   0.63 0.66 &$-$0.50 0.38 &  2   \nl
1443$+$0141 & \nodata  &   0.80 0.33 &   0.29 0.27 &  2   \nl
2201$-$1834 & \nodata  &   0.16 0.07 &   0.00 0.12 &  2   \nl
2350$-$0045 & \nodata  &$-$0.58 0.53 &   0.09 0.30 &  1   \nl
\enddata

\tablenotetext{a}{Wavelength bands $1=$1600-1840\A and $2=$1960-2260\A, rest.}

\tablecomments{Stokes $Q$ and $U$ are listed for objects undetected in 
polarized flux ($<3 \sigma$).}


\end{deluxetable}

\begin{deluxetable}{llcccc}
\tablecaption{BAL trough polarization}
\tablehead{\colhead{IAU (B1950)} & \colhead{Name} & \colhead{$v_{max}$} & 
           \colhead{$Q'\% ~\pm$} & \colhead{$U'\% ~\pm$} & 
           \colhead{dPA$ ~\pm$}}

\startdata
 0019$+$0107  & UM 232   & $-$9500 &  4.8 0.6 &   0.5 0.7 &    3  3  \nl
 0043$+$0048  & UM 275   & $-$3700 &  4.4 0.9 \tablenotemark{a}&   1.7 0.8 \tablenotemark{a} & \nodata \tablenotemark{a} \nl
 0059$-$2735  & \nodata  & $-$2500 &  4.3 1.0 \tablenotemark{b}&$-$1.5 1.1 \tablenotemark{b}&$-$10 8 \tablenotemark{b} \nl
 0105$-$265   & \nodata  &$-$12400 & 10.5 1.6 &$-$0.9 1.7 & $-$2  5  \nl
 0137$-$0153  & UM 356   & $-$5600 &  6.5 2.1 &   2.3 2.2 &   10  6  \nl
 0146$+$0142  & UM 141   &$-$16300 &  4.3 0.7 &$-$1.2 0.7 & $-$8  5  \nl
 0226$-$1024  & \nodata  &$-$10500 &  7.3 0.8 &   1.9 0.4 &   7.3 0.6\nl
 0842$+$3431  & CSO 203  &$-$10500 &  2.9 0.4 &$-$1.0 0.4 &$-$10  5  \nl
 0856$+$1714  & \nodata  & $-$8500 &  8.0 3.6 &   1.3 3.5 &    5 10  \nl
 0903$+$1734  & \nodata  &$-$12400 &  5.2 1.3 &   2.7 1.3 &   14  3  \nl
 0932$+$5006  & \nodata  &$-$14300 &  2.7 0.7 &$-$0.2 0.8 & $-$2  9  \nl
 1212$+$1445  & \nodata  &$-$11400 &  4.3 1.0 &$-$2.6 1.1 &$-$16  8  \nl
 1232$+$1325  & \nodata  &$-$10453 & 12.2 2.6 &   1.2 2.7 &    3  5  \nl
 1235$+$0857  & \nodata  & $-$4600 &  3.3 0.8 &   0.5 0.8 &    4  6  \nl
 1246$-$0542  & \nodata  &$-$15300 &  6.3 0.2 &$-$1.3 0.2 & $-$6  1  \nl
 1331$-$0108  & UM 587   & $-$3700 &  9.0 7.0 &$-$8.0 6.5 &$-$21 22  \nl
 1333$+$2840  & RS 23    & $-$4600 &  8.5 0.9 &   1.3 0.9 &    4  3  \nl
 1413$+$1143  & Clover   & $-$5600 & 11.5 4.5 &  12.3 5.5 &   23  9 \nl
 1524$+$5147  & CSO 755  &$-$12400 &  6.0 0.2 &$-$1.3 0.2 & $-$6  1 \nl
 2225$-$0534  & PHL 5200 & $-$8500 & 10.1 1.1 &$-$1.5 1.1 & $-$4  3 \nl
\enddata

\tablenotetext{a}{For UM 275, ($Q$,$U$) are given instead of 
  ($Q^{\prime}$,$U^{\prime}$), and dPA is undefined because the continuum 
  is unpolarized.}
\tablenotetext{b}{0059$-$2735 trough polarization measured in Mg II BAL.}
\tablecomments
  {The peak trough polarization $Q^{\prime}$ is measured in the C IV BAL 
  (1400-1549\A). $U^{\prime}$ is listed for the velocity $v_{max}$ (km/s) 
  where $Q^{\prime}$ peaks. $v_{max}$ is measured with respect to the peak of 
  the C IV BEL. dPA is the difference PA(trough)$-$PA(continuum), measured 
  at the polarization peak. The formal uncertainties are calculated from 
  photon statistics.}

\clearpage

\end{deluxetable}
  
\begin{deluxetable}{llcccccc}
\tablecaption{Interstellar polarization}
\tablehead{\colhead{IAU (B1950)} & \colhead{Name} & 
           \colhead{$b$\tablenotemark{a}} & 
           \colhead{$P_{max}\%$\tablenotemark{b}} & 
           \colhead{$P_{s}\%$\tablenotemark{c}} & 
           \colhead{PA$_s$\tablenotemark{c}} &
           \colhead{$d$(kpc)\tablenotemark{d}} & 
           \colhead{$d\theta$\tablenotemark{e}}}
\startdata
 0019$+$0107  & UM 232   &$-$60.6 & 0.11 & 0.11 &  96.6 & 0.100 & 0.97 \nl
 0021$-$0213  &          &$-$64.0 & 0.18 & 0.25 & 113.6 & 0.087 & 0.22 \nl
 0025$-$0151  &          &$-$63.8 & 0.18 & 0.25 & 113.6 & 0.087 & 0.96 \nl
 0043$+$0048  & UM 275   &$-$61.8 & 0.02 & 0.29 &  78.4 & 0.229 & 3.02 \nl
 0059$-$2735  &          &$-$87.6 & 0.18 & 0.10 & 133.3 & 0.115 & 2.26 \nl
 0105$-$265   &          &$-$86.2 & 0.05 & 0.10 & 133.3 & 0.115 & 3.90 \nl
 0119$+$0310  &          &$-$58.6 & 0.09 & 0.08 & 156.6 & 0.229 & 0.67 \nl
 0137$-$0153  & UM 356   &$-$61.9 & 0.11 & 0.08 & 128.5 & 0.182 & 1.52 \nl
 0145$+$0416  & UM 139   &$-$55.5 & 0.25 & 0.30 & 128.4 & 0.316 & 1.60 \nl
 0146$+$0142  & UM 141   &$-$57.7 & 0.00 & 0.30 & 128.4 & 0.316 & 3.31 \nl
 0226$-$1024  &          &$-$61.6 & 0.02 & .... & ..... & ..... & .... \nl
 07598$+$6508 & IRAS     &   32.1 & 0.32 & .... & ..... & ..... & .... \nl 
 0842$+$3431  & CSO 203  &   37.5 & 0.18 & .... & ..... & ..... & .... \nl 
 0846$+$1540  &          &   32.9 & 0.20 & 0.02 &  35.0 & 0.063 & 2.82 \nl
 0856$+$1714  &          &   35.7 & 0.11 & 0.02 &  35.0 & 0.063 & 3.78 \nl
 0903$+$1734  &          &   37.5 & 0.16 & 0.03 &  86.0 & 0.016 & 2.81 \nl
 0932$+$5006  &          &   46.5 & 0.07 & 0.01 & 115.0 & 0.017 & 1.94 \nl
 1011$+$0906  &          &   48.5 & 0.07 & 0.06 &  82.7 & 0.066 & 1.80 \nl
 1212$+$1445  &          &   74.7 & 0.27 & 0.03 & 116.0 & 0.052 & 0.58 \nl
 1231$+$1320  &          &   75.4 & 0.11 & 0.05 &  74.0 & ..... & 2.80 \nl
 1232$+$1325  &          &   75.5 & 0.14 & 0.05 &  74.0 & ..... & 2.91 \tablebreak
 1235$+$0857  &          &   71.3 & 0.00 & 0.02 &  39.0 & 0.060 & 1.88 \nl
 1235$+$1453  &          &   77.1 & 0.11 & 0.22 &  76.0 & ..... & 2.47 \nl
 1243$+$0121  &          &   63.9 & 0.00 & 0.22 &  76.0 & ..... & 2.47 \nl
 1246$-$0542  &          &   56.9 & 0.09 & 0.26 &  45.9 & 0.380 & 2.25 \nl
 1331$-$0108  & UM 587   &   59.7 & 0.11 & 0.04 &   9.0 & 0.029 & 0.81 \nl
 1333$+$2840  & RS 23    &   80.0 & 0.00 & 0.08 & 104.0 & ..... & 1.11 \nl
 1413$+$1143  & Clover   &   64.8 & 0.00 & 0.10 &  57.0 & 0.105 & 1.41 \nl
 1442$-$0011  &          &   51.2 & 0.29 & 0.15 &  91.0 & 0.035 & 2.33 \nl
 1443$+$0141  &          &   52.5 & 0.27 & 0.05 &  78.0 & 0.036 & 0.44 \nl
 1524$+$5147  & CSO 755  &   52.1 & 0.07 & .... & ..... & ..... & .... \nl
 1700$+$5153  & PG       &   37.8 & 0.05 & .... & ..... & ..... & .... \nl
 2154$-$2005  &          &$-$49.6 & 0.09 & 0.09 &   6.5 & 0.229 & 1.42 \nl
 2201$-$1834  &          &$-$50.6 & 0.16 & 0.08 & 175.9 & 0.437 & 1.29 \nl
 2225$-$0534  & PHL 5200 &$-$49.6 & 0.38 & 0.44 & 160.5 & 0.087 & 2.23 \nl
 2350$-$0045A &          &$-$59.9 & 0.27 & 0.10 & 128.9 & 0.060 & 2.59 \nl
\enddata
\tablenotetext{a}{Galactic latitude}
\tablenotetext{b}{Upper limit to the interstellar polarization}
\tablenotetext{c}{Polarization $P_s$ and electric vector PA$_s$ of a nearby 
                  interstellar probe star}
\tablenotetext{d}{Probe star distance from the observer}
\tablenotetext{e}{Probe star angular distance d$\theta$ ($\arcdeg$) from the 
                  BAL QSO}

\end{deluxetable}


\clearpage

\begin{thebibliography}{36}

\bibitem[Antonucci et~al. 1996]{agg96}
Antonucci, R., Geller, R., Goodrich, R.~W., \& Miller, J.~S. 
  1996, ApJ, 472, 502

\bibitem[Barlow 1993]{b93}
Barlow, T.~A. 1993.
\newblock Time Variability of Broad Absorption-Line QSOs.
\newblock Ph.D. thesis, University of California, San Diego

\bibitem[Barlow et~al. 1992]{bjb92}
Barlow, T.~A., Junkkarinen, V.~T., Burbidge, E.~M., Weymann, R.~J., et~al.
  1992, ApJ, 397, 81

\bibitem[Becker et~al. 1997]{bgh97}
Becker, R. H., Gregg, M. D., Hook, I. M., McMahon, R. G., White, R. L.,
\& Helfand, D. J. 1997, ApJ, 479, L93

\bibitem[Beland et~al. 1988]{bbd88}
Beland, S., Boulade, O., \& Davidge, T. 1988.
\newblock Tech. Rep.~19, CFHT.
\newblock 16

\bibitem[Boroson \& Meyers 1992]{bm92}
Boroson, T.~A. \& Meyers, K.~A. 1992, ApJ, 397, 442

\bibitem[Brotherton et~al. 1998]{bwd98}
Brotherton, M.~S., Wills, B.~J., Dey, A., Van Breugel, W., \& Antonucci, R.
1998, ApJ, 501, 110

\bibitem[Brotherton et~al. 1997]{btb97}
Brotherton, M.~S., Tran, H.~D., Van Breugel, W., Dey, A., \& Antonucci, R.
1997, ApJL, 487, 113

\bibitem[Burstein \& Heiles 1984]{bh84}
Burstein \& Heiles 1984, ApJS, 54, 33

\bibitem[Canalizo \& Stockton 1997]{cs97}
Canalizo, G. \& Stockton, A. 1997, ApJL, 480, 5

\bibitem[Chandrasekhar 1960]{c60}
Chandrasekhar, S. 1960.
\newblock Radiative Transfer.
\newblock New York: Dover

\bibitem[Clarke et~al. 1983]{cs83}
Clarke, D., Stewart, B.~G., Schwarz, H.~E., \& Brooks, A. 1983, A\&A, 126, 260

\bibitem[Cohen et~al. 1999]{cot99}
Cohen, M.~H., Ogle, P.~M., Tran, H.~D., Goodrich, R.~W., \&Miller, J.~S. 1999.
\newblock Submitted to AJ.

\bibitem[Cohen et~al. 1995]{cot95}
Cohen, M.~H., Ogle, P.~M., Tran, H.~D., Vermeulen, R.~C., \& Miller, J.~S.
  1995, ApJ, 448, L77

\bibitem[Cohen et~al. 1997]{cvo97}
Cohen, M.~H., Vermeulen, R.~C., Ogle, P.~M., Tran, H.~D., \& Goodrich, R.~W.
  1997, ApJ, 484, 193

\bibitem[De Breuck et~al. 1998]{dbt98}
De Breuck, C., Brotherton, M.~S, Tran, H.~D., Van Breugel, W., \&
Rottgering, H.~J.~A. 1998, AJ, 116, 13

\bibitem[Francis et~al. 1992]{fhf92}
Francis, P. J., Hewett, P. C., Foltz, C. B., \& Chaffee, F. H. 1992,
ApJ 398, 476

\bibitem[Glenn et~al. 1994]{gsf94}
Glenn, J., Schmidt, G., \& Foltz, C. 1994, ApJ, 434, L47

\bibitem[Goodrich 1997]{g97}
Goodrich, R.~W. 1997, ApJ, 474, 606

\bibitem[Goodrich et al. 1996]{gmm96}
Goodrich, R.~W., Miller, J.~S., Martel, A., Cohen, M.~H., Tran, H.~D.,
Ogle, P.~M., and Vermeulen, R.~C. 1996, ApJ, 456, L9

\bibitem[Goodrich \& Miller 1995]{gm95}
Goodrich, R.~W. \& Miller, J.~S. 1995, ApJ, 448, L73

\bibitem[Goodrich \& Miller 1994]{gm94}
Goodrich, R.~W., \& Miller, J.~S. 1994, ApJ, 434, 82

\bibitem[Goodrich 1991]{g91}
Goodrich, R.~W. 1991, PASP, 107, 1314

\bibitem[Goodrich \& Miller 1988]{gm88}
Goodrich, R.~W. \& Miller, J.~S. 1988, ApJ, 331, 332

\bibitem[Goodrich 1986]{g86}
Goodrich, R.~W. 1986, ApJ, 311, 882

\bibitem[Hamman 1999]{h99}
Hamann, F. 1999, ApJ (in press)

\bibitem[Hamman, Korista, \& Morris 1993]{hkm93}
Hamann, F., Korista, K. T., \& Morris, S. L. 1993. ApJ 415, 541

\bibitem[Hayes 1971]{h71}
Hayes, D.~S. 1971, ApJ, 159, 165

\bibitem[Hazard et~al. 1987]{hmw87}
Hazard, C., McMahon, R.~G., Webb, J.~K., \& Morton, D.~C. 1987, 
  ApJ, 323, 263

\bibitem[Hines \& Wills 1995]{hw95}
Hines, D.~C. \& Wills, B.~J. 1995, ApJ, 448, L69

\bibitem[Hines et al. 1999]{hlt99}
Hines, D.~C., Low, F.~J., Thompson, R.~I., Weymann, R.~J., \&
Storrie-Lombardi, L.~J. 1999, ApJ (in press)

\bibitem[Kellermann et~al. 1994]{kss94}
Kellermann, K.~I., Sramek, R., Schmidt, M., Green, R.~F., \& Shaffer, D.~B.
  1994, AJ, 108, 1162

\bibitem[Kneib et~al. 1998]{kam98}
Kneib, J.-P., Alloin, D., Mellier, Y., Guilloteau, S., Barvainis, R.,
\& Antonucci, R. 1998 A\&A, 329, 827

\bibitem[Korista \& Ferland 1998]{kf98}
Korista, K. \& Ferland, G. 1998, ApJ 495, 672

\bibitem[Lightman \& Shapiro 1975]{ls75}
Lightman, A.~P. \& Shapiro, S.~L. 1975, ApJ, 198, L73

\bibitem[Magain et~al. 1988]{mss88}
Magain, P., Surdej, J., Swings, J.~P., Borgeest, U., et~al. 1988, Nature, 334,
  325

\bibitem[Martel 1996]{m96}
Martel, A.~R. 1996.
\newblock Spectropolarimetry of High-Polarization Seyfert 1 Galaxies.
\newblock Ph.D. thesis, University of California, Santa Cruz

\bibitem[Matthewson \& Ford 1970]{mf70}
Matthewson \& Ford 1970, Mem. RAS, 74, 139

\bibitem[Miller, Robinson, \& Goodrich 1988]{mrg88}
Miller, J.~S., Robinson, L. B., \& Goodrich, R. W. 1988, in Instrumentation
for Ground-based Astronomy, ed. L. B. Robinson (New York:Springer), 157

\bibitem[Naghizadeh-Khouei \& Clarke 1993]{nc93}
Naghizadeh-Khouei, J. \& Clarke, D. 1993, A\&A, 274, 968

\bibitem[Ogle 1999b]{o99b}
Ogle, P.~M. 1999b. Paper II, in preparation.

\bibitem[Ogle 1999c]{o99c}
Ogle, P.~M. 1999c. Paper III, in preparation.

\bibitem[Ogle 1998]{o98} 
Ogle, P.~M. 1998. Polarization and Structure of Broad Absorption Line
Quasi-Stellar Objects. Ph.D. Thesis, California Institute of Technology. 

\bibitem[Ogle et al. 1997]{ocm97}
Ogle, P.~M., Cohen, M.~H., Miller, J.~S., Tran, H. D., Fosbury, R.~A.~E.,
\& Goodrich, R.~W. 1997, ApJ, 482, L37

\bibitem[Ogle 1997]{o97}
Ogle, P.~M. 1997. ASP Conference Series, Vol. 128, Mass Ejection from AGN,
p. 78, Eds. N. Arav, I. Shlosman, and R. J. Weymann.

\bibitem[Oke 1990]{o90}
Oke, J.~B. 1990, AJ, 99, 1621

\bibitem[Oke et~al. 1995]{o95}
Oke, J.~B., Cohen, J.~G., Carr, J., Cromer, J., et~al. 1995, PASP, 107, 375

\bibitem[Oke \& Gunn 1982]{og82}
Oke, J.~B. \& Gunn, J.~E. 1982, PASP, 94, 586

\bibitem[Pettini \& Boksenberg 1985]{pb85}
Pettini, M. \& Boksenberg, A. 1985, ApJ, 294, L73

\bibitem[Pettini et~al. 1997]{pks97}
Pettini, M., King, D.~L., Smith, L.~J., \& Hunstead, R.~W. 1997, ApJ, 478, 536

\bibitem[Putney \& Cohen 1996]{pc96}
Putney, A. \& Cohen, M.~H. 1996.
\newblock Second Order Light Effects in LRIS.
\newblock Tech. rep., California Institute of Technology

\bibitem[Rudy \& Schmidt 1988]{rs88}
Rudy, R.~J. \& Schmidt, G.~D. 1988, ApJ, 331, 325

\bibitem[Schmidt \& Hines 1999]{sh99}
Schmidt, G.~D. \& Hines, D.~C. 1999, ApJ, 512, 125.

\bibitem[Schmidt et~al. 1997]{shs97}
Schmidt, G.~D., Hines, D.~C., \& Smith, P.~S. 1997.
\newblock In Mass Ejection from Active Galactic Nuclei, eds. N.~Arav,
  I.~Shlosman, \& R.~J. Weymann, vol. 128 of ASP Conference Series,  305

\bibitem[Schmidt et~al. 1992]{sel92}
Schmidt, G.~D., Elston, R., \& Lupie, O.~L. 1992, AJ, 104, 1563

\bibitem[Serkowski et~al. 1975]{smf75}
Serkowski, K., Mathewson, D.~S., \& Ford, V. 1975, ApJ, 196, 261

\bibitem[Simmons \& Stewart 1985]{ss85}
Simmons, J. F.~L. \& Stewart, B.~G. 1985, A\&A, 172, L11

\bibitem[Sowinski et~al. (1997)]{ssh97}
Sowinski, L.~G., Schmidt, G.~D., \& Hines, D.~C. 1997.
\newblock In Mass Ejection from Active Galactic Nuclei, eds. N.~Arav,
  I.~Shlosman, \& R.~J. Weymann, vol. 128 of ASP Conference Series,  305

\bibitem[Stockman et~al. 1984]{sma84}
Stockman, H.~S., Moore, R.~L., \& Angel, J. R.~P. 1984, ApJ, 279, 485

\bibitem[Stockman et~al. 1981]{sah81}
Stockman, H.~S., Angel, J. R.~P., \& Hier, R.~G. 1981, ApJ, 243, 404

\bibitem[Stockton et~al. 1998]{scc98}
Stockton, A., Canilizo, G., \& Close, L.~M. 1998, ApJ
\newblock In press

\bibitem[Sun \& Malkan 1989]{sm89}
Sun, W.-H. \& Malkan, M. 1989, ApJ, 346

\bibitem[Tran et~al. 1998]{tco98}
Tran, H.~D., Cohen, M.~H., Ogle, P.~M., Goodrich, R.~W., \&
di Serego Alighieri, S. 1998, ApJ, 500, 660.

\bibitem[Tran, Cohen, \& Goodrich 1995]{tcg95}
Tran, H.~D., Cohen, M.~H., \& Goodrich, R.~W. 1995, AJ, 110, 2597

\bibitem[Tran 1995]{t95}
Tran, H.~D. 1995, ApJ, 440, 565

\bibitem[Wampler, Chugai, \& Petitjean 1995]{wcp95}
Wampler, E. J., Chugai, N. N., \& Petitjean, P. 1995, ApJ, 443, 586

\bibitem[Weymann et~al. 1991]{wmf91}
Weymann, R.~J., Morris, S.~L., Foltz, C.~B., \& Hewett, P.~C. 1991, ApJ, 373,
  23, WMF

\bibitem[Webb et~al. 1993]{wms93}
Webb, W., Malkan, M., Schmidt, G., \& Impey, C. 1993, ApJ, 419, 494

\bibitem[Wills \& Hines 1997]{wh97}
Wills, B.~J., \& Hines, D.~C. 1997.
\newblock In Mass Ejection from Active Galactic Nuclei, eds. N.~Arav,
  I.~Shlosman, \& R.~J. Weymann, vol. 128 of ASP Conference Series, 99

\bibitem[Wills et al. 1992]{wwe92}
Wills, B.~J., Wills, D., Evans, N.~J., II, Natta, A., Thompson, K. L.,
Breger, M., and Sitko, M. L. 1992, ApJ 400, 96

\bibitem[Wills, Netzer, \& Wills 1985]{wnw85}
Wills, B.~J., Netzer, H., \& Wills, D. 1985, ApJ, 288, 94

\end{thebibliography}
\end{document}